\documentclass[twocolumn, twocolappendix]{aastex63}
\usepackage{tcolorbox} 
\usepackage[italicdiff]{physics} 
\usepackage[utf8]{inputenc}
\usepackage{amssymb,natbib,graphicx,color,ctable, amsmath,array,enumitem,kantlipsum}
\usepackage{hyperref}
\usepackage{scalerel}
\usepackage{xspace}
\hypersetup{
    colorlinks=true,
    linkcolor=blue,
    citecolor=blue,
    filecolor=blue,
    urlcolor=blue
}

\usepackage[flushleft]{threeparttable}
\usepackage{lineno}
\newcommand\hi{H\,\protect\scaleto{$I$}{1.2ex}}
\newcommand\hii{H\,\protect\scaleto{$II$}{1.2ex}}
\newcommand\hiv{H\protect\scaleto{$I$}{1.2ex}}

\newcommand\ehiv{\text{\hiv}}
\newcommand\skirt{\textsc{Skirt}}

\defcitealias{Stern2021}{S21}

\begin{document}
\title{Molecular Hydrogen in High-redshift Damped Lyman-$\alpha$ Absorbers}
\author[0009-0004-2434-8682]{Alon Gurman}
\affiliation{School of Physics \& Astronomy, Tel Aviv University, Ramat Aviv 69978, Israel}
\author[0000-0001-5065-9530]{Amiel Sternberg}
\affiliation{School of Physics \& Astronomy, Tel Aviv University, Ramat Aviv 69978, Israel}
\affiliation{Center for Computational Astrophysics, Flatiron Institute, 162 5th Avenue, New York, NY 10010, USA}
\affiliation{Max-Planck-Institut f\"{u}r Extraterrestrische Physik, Giessenbachstrasse 1, D-85748 Garching, Germany}
\author[0000-0002-0404-003X]{Shmuel Bialy}
\affiliation{Physics Department, Technion, Haifa, 3200003, Israel}
\author[0000-0001-8855-6107]{Rachel K. Cochrane}
\affiliation{Jodrell Bank Centre for Astrophysics, University of Manchester, Oxford Road, Manchester M13 9PL, UK}
\affiliation{Department of Astronomy, Columbia University, New York, NY 10027, USA}
\affiliation{Center for Computational Astrophysics, Flatiron Institute, 162 5th Avenue, New York, NY 10010, USA}
\author[0000-0002-7541-9565]{Jonathan Stern}
\affiliation{School of Physics \& Astronomy, Tel Aviv University, Ramat Aviv 69978, Israel}

\correspondingauthor{Alon Gurman}
\email{alongurman@gmail.com}

\begin{abstract}
Simulations predict that circumgalactic hydrogen gas surrounding massive ($M_{\rm{halo}}^{z=1}=10^{12}-10^{13}\ M_{\odot}$) galaxies at $z\sim4$ may be predominantly neutral, and could produce damped Ly$\alpha$ absorbers (DLAs) along sight-lines to background quasars \citep{Stern2021}. A circumgalactic medium (CGM) origin for DLAs naturally explains \hi{} absorption-selected galaxy detections at $z\sim4$ \citep{Neeleman2017, Neeleman2019, Neeleman2025} with physical separations much greater than the likely extents of the galaxy disks. These \hi{} column densities observed in $z\sim 4$ DLA  are large and comparable to interstellar (ISM) gas columns at which substantial molecular hydrogen (H$_2$) abundances occur. We therefore investigate the possible molecular content of high-redshift CGM gas, and its potential detectability via (rest-frame) far-ultraviolet (UV) absorption line studies.  For this purpose we develop an analytic sub-grid model for \hiv{}-to-H$_2$ transitions and incorporate the model with zoom-in FIRE-2 simulations of evolving high-$z$ galaxies. We include dust absorption and scattering computations for the transfer of photodissociating Lyman-Werner (LW) band radiation. We find that the typical extents of detectable H$_2$ sightlines are $\approx 0.1\, R_{\rm vir}$, independent of redshift from $z=2.5$ to 5. We argue that a CGM origin for DLAs naturally explains the low detection rates of H$_2$ in DLA observations, as the low CGM densities and relatively strong far-UV fields lead to molecular fractions much lower than observed in the ISM at comparable \hi{} columns.
\end{abstract}

\keywords{
Circumgalactic medium (1879) -- Astrochemistry (75) -- Hydrodynamical simulations (767) -- Damped Lyamn-alpha systems (349) -- Galaxy evolution (594)
}
\section{Introduction}
Damped Ly$\alpha$ absorbers (DLAs), lines of sight to background quasars with neutral hydrogen (\hi{}) column densities $N_{\ehiv}>2\times 10^{20} \;\rm{cm}^{-2}$, present both open questions and insights into the \hi{} gas content of the universe. Ever-growing samples of DLA detections found in optical surveys across a wide redshift range are now readily available \citep{Prochaska2009, Noterdaeme2012, Ho2020}. In the standard interpretation DLAs trace the \hi{} interstellar medium (ISM) of galaxy disks \citep{Wolfe2005} but many features remain open to study. 

Many absorption-selected detections of DLA galaxy counterparts have been reported in the literature. 
Optical/NIR studies have revealed several galaxy counterparts within $\lesssim10$ kpc of the DLAs \citep{Fynbo2010,Moller2018}. With SINFONI on the VLT, \citet{Peroux2011,Peroux2012,Peroux2016} targeted $z\approx1-2$ DLAs and detectedoptical nebular lines inclduing H$\alpha$ emission. They estimated star formation rates (SFRs) in the range 1.3-3.5 $M_{\odot}\  \rm yr^{-1}$ range with one source displaying an SFR of 17.1 $M_{\odot}\  \rm yr^{-1}$. 
While some of their detections had impact parameters $b\lesssim10$ kpc, consistent with the DLAs being part of the galaxy disks, others with $b>25$ kpc were suggested to be of CGM origin. Ly$\alpha$ emission has also been detected in the center of the Ly$\alpha$ trough of DLAs at impact parameters ranging from $\sim0$ to 39 kpc \citep{Moller2004,Noterdaeme2012b,Krogager2012,Peroux2016}. However, the small sizes of the DLA galaxy counterparts have made the interpretation difficult, with a single ubiquitous picture still lacking.

Of particular interest are a set of DLAs at redshifts $z\sim4$ for which galaxy counterparts were discovered via the [C$\,${\small II}] 158 $\mu$m fine-structure emission line \citep{Neeleman2017, Neeleman2019,Neeleman2025} with impact parameters often greater than 20 kpc and as large as 60 kpc, and velocity offsets consistently below $100\ \rm km\ s^{-1}$. Using measurements of dust continuum far-infrared emission, they deduced star formation rates ranging from less than 10 to 110 $M_{\odot}\ \rm yr ^{-1}$, with typical values of order $10\ M_{\odot}\ \rm yr^{-1}$. In addition, over 100  Lyman-$\alpha$ emitter counterparts to DLAs at $z\approx3-4$ with impact parameters $\gtrsim100 \ \rm kpc$ have been observed \citep{Mackenzie2019,Lofthouse2023}. The physical separations of the aforementioned DLAs from the [C~II] and Lyman-$\alpha$ counterparts are too large to be plausibly associated with the galaxy gaseous disks.

\citet{Stern2021} (hereafter \citetalias{Stern2021}) suggested that these DLAs are produced within the more extended circumgalactic medium (CGM) surrounding the high redshift galaxies.
\citetalias{Stern2021} used analytic arguments and the FIRE-2 simulations \citep{Hopkins2018} to show that at $z\sim 4$ CGM gas is likely primarily neutral out to significant fractions of the halo virial radii $R_{\rm vir}$. This is because the higher halo baryon densities at high-$z$ enable efficient cooling and shielding from any background ionizing radiation \citep[see also][]{Theuns2021,Gurvich2023}. 

In this paper, we address the question of whether detectable quantities of molecular hydrogen (H$_2$) might be produced in the neutral CGM of high-$z$ galaxies. We are motivated by several observational absorption line studies of H$_2$ in DLAs that have yielded either detections or upper limits \citep{Ledoux2003,Noterdaeme2008,Noterdaeme2015}. 

In the far-UV irradiated interstellar medium (ISM) of the Milky Way and  nearby galaxies,
H$_2$ forms efficiently in dense cold clouds where the combination of dust opacity  and H$_2$ self-shielding enables conversion from  \hi{} to H$_2$ \citep{Bigiel2008,Leroy2008,Schruba2011,Sternberg2014}. In the ISM of the Galaxy the molecular hydrogen column density $N_{\rm H_2}$ rises steeply as a function of $N_{\ehiv}$ when $N_{\ehiv}\approx5\times 10^{20}\ \rm cm^{-2}$, displaying values  $f_{\rm H_2}\equiv2N_{\rm H_2}/\left(N_{\ehiv}+2N_{\rm H_2}\right)> 0.1$. This has been determined through UV absorption studies of lines of sight towards stars or AGN and emission studies of molecular clouds \citep{Savage1977, Allen2004,Gillmon2006a,Gillmon2006b,Lee2012,Lee2015,Bialy2015c,Rachford2009,VanDePutte2023}. 

In contrast, observations of H$_2$ in DLAs paint a different picture, where it is common to observe $f_{\rm H_2}\ll 0.01$ even when $N_{\ehiv{}}\gtrsim 10^{21} \ \rm cm^{-2}$. \cite{Ledoux2003}, observed 33 DLAs with the Ultraviolet and Visual Echelle Spectrograph (UVES) on the Very Large Telescope (VLT), with the goal of detecting H$_2$ and estimating the molecular column densities. Their study, as well as additional observations by \cite{Noterdaeme2008,Noterdaeme2014,Noterdaeme2015}, span a redshift range of $z=1.8{-}4.2$ and \hi{} column density range of $\log \left(N_{\ehiv{}}/\rm{cm}^{-2}\right)=19.35-22.4$. Their detection limit was $N_{\rm H_2,detec}= 2\times10^{14} \ \rm cm^{-2}$. Most sightlines yielded H$_2$ non-detections, implying H$_2$ mass fractions $f_{\rm H_2}<10^{-4}$ even in DLAs with $N_{\ehiv{}}>10^{21}\ \rm{cm}^{-2}$. Only four sightlines yielded H$_2$ detections with $f_{\rm H_2}\gtrsim0.01$. H$_2$ detections are more common than upper-limits only for $\log \left(N_{\ehiv{}}/\mathrm{cm ^{-2}}\right)>21.5$.

In this paper, we ask whether these observations are consistent with DLA and H$_2$ formation in neutral CGM gas. To address this question, we utilize the FIRE-2 simulations and a subgrid model for H$_2$ formation and destruction. We apply our subgrid model to the simulation outputs in post-processing.   We also include dust-radiative transfer calculations to account for attenuation of stellar and metagalactic photodissociating Lyman-Werner ($11.2-13.6$ eV; LW) band radiation. While previous works have modeled the H$_2$ content of galaxies in cosmological simulations \citep{Gnedin2009,Christensen2012,Diemer2018,Diemer2019a,Gebek2023}, in this work we focus on extending our analysis to the CGM of simulated galaxies. 

We describe our models in Section \ref{sec: model ingredients}, and present our results in Section \ref{sec: results}. In Section \ref{sec: observations} we compare our results to observational data and suggest avenues for future study. We summarize in Section \ref{sec: summary}.

\section{Model Ingredients}
\label{sec: model ingredients}

In this section we describe our subgrid \hiv{}-to-H$_2$ transition model and its application. The model is based on the 1-dimensional (1D) analytic theory presented in \citet{Bialy2016a}, and allows us to compute the H$_2$ column density for a given set of cloud parameters. These include the hydrogen nucleus density $n$, the LW band intensity $I_{\rm LW}$ relative to the \citet{Draine1978} field, and the normalized dust abundance $Z^\prime_d$. We apply our model in post-processing of the FIRE-2 hydrodynamical simulations by treating the neutral gas in each FIRE-2 gas cell as a 1D cloud, each with a given total hydrogen gas density and boundary LW band flux. We determine the LW band fluxes with additional post-processing of the FIRE-2 outputs using the radiative transfer code \skirt{} \citep{Baes2011,Camps2015}. 

\subsection{FIRE-2 Simulations}

The Feedback in Realistic Environments \citep[FIRE;][]{Hopkins2014,Hopkins2018,Hopkins2023} project provides a suite of high-resolution galaxy evolution simulations in a cosmological context. The FIRE simulations were developed to explore the role of feedback in galaxy formation and evolution. FIRE uses the zoom-in technique, where galaxy formation is simulated at high-resolution while initial and boundary conditions are determined by a lower-resolution, cosmological dark matter only simulation. 

In this work we use results from the second iteration FIRE-2 simulations to estimate H$_2$ abundances using our post-processing methodology. The FIRE-2 computations are based on the multi-method gravity and hydrodynamics code \textsc{Gizmo} \citep{Hopkins2015} in its meshless finite-mass mode. Star formation takes place above a hydrogen density threshold of $n_{\rm H}>1000\ \rm cm ^{-3}$. Sub-grid models are used to account for the effects of supernovae (SNe), stellar winds, metal deposition, radiation pressure, photoionization heating, and dust photoelectric heating. Gas cooling processes include metal line cooling, free-free emission, and Compton scattering with the cosmic microwave background.

The neutral atomic (\hi{}) hydrogen abundance is determined by a balance between recombination and photoionization from both stellar sources and a metagalactic ultraviolet (UV) background \citep{Faucher-Giguere2009}. The computation of the local radiation field includes shielding by neighbouring gas cells and cells near the radiation sources. For more details on the implementation of the different physical mechanisms in the FIRE-2 simulations, see \citet{Hopkins2018}.

In this paper we analyze the MassiveFIRE suite of simulations presented in \citet{AnglesAlcazar2017}. The halo selection was described in \citet{Feldmann2016,Feldmann2017}. The snapshot data has been made public and is described in \citet{Wetzel2023}. They all share a baryonic mass resolution of 33000 $M_{\odot}$ and are run down to redshift $z=1$. We use snapshots in the redshift range $z=2.5-5$ with 0.5 increments. A brief summary of the galaxy properties is given in Table \ref{table: galaxy props}. These halos were selected for their high masses, with stellar masses of few $10^{11}\ M_{\odot}$ by $z=1$. The mean star formation rate and stellar mass at $z=4$ are 9.59 $M_{\odot}\  \rm yr^{-1}$ and $6.13\times 10^9 \ M_{\odot}$, respectively.

\begin{table}
\hspace*{-1.5cm}
\centering
\begin{threeparttable}

\centering 
\begin{tabular}{l l l l}
\hline \hline 
\\ [-1.5ex] Name & $M_{\rm halo} ^{z=1}\ \left[M_{\odot} \right]$ & $M_{\star} ^{z=1}\ \left[M_{\odot} \right]$ & $z$  \\ [0.5ex] 
\hline

\\[-1ex] A1 & $0.4\times10^{13}$ & $2.8\times 10^{11}$ & $2.5-5$
\\[0.5ex] A2 & $0.5\times10^{13}$ & $2.7\times 10^{11}$ & $3-5$
\\[0.5ex] A4 & $0.8\times10^{13}$ & $5.1\times 10^{11}$ & $2.5-5$
\\[0.5ex] A8 & $1.3\times10^{13}$ & $8.0\times 10^{11}$ & $2.5-5$
\\[0.5ex] 
\hline

\end{tabular}
\end{threeparttable}
\caption{Galaxy properties for the MassiveFIRE suite \citep{AnglesAlcazar2017}.}
\label{table: galaxy props} 
\end{table}

\subsection{Radiative Transfer with SKIRT}

An important factor in setting the H$_2$ abundance is the spatial distribution of the LW band flux. To this end, we follow the methodology presented in \citet{Cochrane2019, Cochrane2023}, in which the Monte Carlo radiative transfer code \skirt{} is utilized \citep{Baes2011,Camps2015}. \skirt{} propagates photons from radiation sources through an input spatial distribution of dust, and computes the effects of dust absorption, scattering, and re-emission of absorbed light. The light crossing times are shorter than the relevant dynamical times, justifying this post-processing approach. 

We extract gas and star particles from the MassiveFIRE snapshot data within the virial radius, $R_{\rm vir}$, of each halo. We assume that gas particles with temperature $T<10^6\;\rm K$ have an ISM-like dust-to-metal ratio of 0.4 \citep{Dwek1998,James2002}, and use the metallicity calculated in the simulation. For gas particles with $T>10^6\ \rm K$ we assume that dust is absent due to grain sputtering \citep{Draine1979, Tielens1994a}.
Our assumption that the CGM is dusty could be invalid if the cool CGM 
condensed from hot $T\sim 10^6\ \rm K$ gas. Our model would then overestimate the effects of far-UV dust absorption, leading to overestimated H$_2$ fractions. However, there is observational evidence for a dust abundance in the CGM that is similar to that of the host galaxy \citep{Menard2010}. A more self-consistent model could potentially include an explicit treatment of dust evolution \citep[see, e.g.,][]{Choban2022}, but such treatment is beyond the scope of this work.

For the dust radiative transfer we use a \citet{Weingartner2001b} Milky Way prescription for a mixture of graphite, silicate, and polycyclic aromatic hydrocarbon (PAH) grains. Star particles are assigned spectral energy distributions (SEDs) according to their ages and metallicities following \citet{Bruzual2003}. We use an octree dust grid, in which cell sizes are adjusted according to the dust density distribution. Since we are interested in H$_2$ dissociation, we specify the input wavelength for \skirt{} to be 10 equally spaced discrete points in the (LW) band, i.e., from 912 to 1107 \AA. 

\skirt{} outputs the energy density in each energy bin in each octree cell. Next, we convert the energy densities into photon fluxes, and add them up to obtain the total LW band photon flux $F_{\rm LW}$. Finally, we assign every gas particle the $F_{\rm LW}$ value of the octree cell in which it is contained.

\subsection{H$_2$ Chemistry}
\label{subsec: H2 model}

In our subgrid model, we compute the H$_2$ fractions by treating the neutral gas in each FIRE-2 gas cell as a two-sided plane parallel 1D slab, with uniform (atomic+molecular) hydrogen nucleus density and dust abundance, with both sides irradiated by the transferred LW band radiation flux. The H$_2$ abundance at each point within the cloud is computed assuming a steady state at which the local  H$_2$ formation and destruction rates are equal. 
We describe the details of our model below.

\subsubsection{H$_2$ Formation}
At sufficiently high metallicity, H$_2$ forms on dust grains out of two H atoms absorbed onto the surface of dust grains

\begin{equation*}
    \rm{H}\;+\;\rm{H}\;+\;\rm{grain}\;\to\;\rm{H}_2\;+\;\rm{grain},
    \label{reac: dust form}
\end{equation*}
either by diffusion or collisions, with the dust grain absorbing the excess energy and the H$_2$ molecule ejected into the gas \citep{Gould1963,Hollenbach1970,Cazaux2004b}. We adopt a simple expression for the dust grain H$_2$ formation rate coefficient, defined as the formation rate per unit density

\begin{equation}
    R_d = 3.0\times10^{-17} \left(\frac{T}{100\ \rm K} \right)^{1/2} Z^{\prime}_d\ \ \ \  \rm cm ^3 \ \rm s^{-1} \ ,
\end{equation}
where $Z^{\prime}_d$ is the dust abundance relative to the galactic value of 0.01, and $T$ is the gas temperature \citep[see][]{Sternberg2014}. We assume $T=100 \ \rm K$ and ignore any temperature dependence in $R_d$. This is justified because we do not compute or resolve the temperature structure in our subgrid model. Including a temperature dependence for $R_d$ would add complexity without making our models more consistent. In addition, as we will describe in Section \ref{sec: subgrid application}, our assumption that $Z^{\prime}_d$ is linearly proportional to the gas metallicity, is another simplification that adds uncertainty to our adopoted values for $R_d$. Conceivably, the dust-to-metals ratio at high redshift and in the CGM could be lower than in the ISM of the Galaxy, which would lower our predictions for the H$_2$ columns.

We ignore H$_2$ formation in the gas phase, a formation channel that is important only at metallicities $\lesssim10^{-3}$ \citep[see, e.g.,][]{Sternberg2021}. We discuss the potential effect of including gas phase H$_2$ formation in Appendix A.

\subsubsection{H$_2$ Destruction}

Destruction of H$_2$ is assumed to take place via LW band photodissociation. The free-space unattenuated photodissociation rate $D_{0, \rm LW}$ is given by 

\begin{equation}
    D_{0, \rm LW}=5.8\times 10^{-11} I_{\rm{LW}} \;\rm{s}^{-1},
\end{equation}
\citep{Sternberg2014}
where $I_{\rm{LW}}$ is the flux in the LW energy band, relative to the \cite{Draine1978} field photon flux of $2.07\times 10^7\;\rm{s}^{-1}\;\rm{cm}^{-2}$. LW band photons are assumed to be absorbed via both dust grain absorption and H$_2$ line self-shielding. 

The photodissociation rate in a cloud irradiated from one side, at a depth with total (\hi{} plus H$_2$) column density $N$, and H$_2$ column density $N_{\rm{H}_2}$, is given by
\begin{equation}
    D_{\rm LW} \left(N \right)=\frac{1}{2} D_{0, \rm LW} e^{-\tau _d} f_{\rm{sh}}(N_{\rm{H}_2})
    \label{eq: D and f_sh}
\end{equation}
\citep{Sternberg2014}. In this expression, $\tau_d$ and $N_{\rm H _2}$ are the dust absorption optical depth in the LW band and the H$_2$ column density at depth $N$, respectively. $f_{\rm{sh}}(N_{\rm{H}_2})$ is the LW band flux attenuation factor due to H$_2$ self-shielding, and is a function of $N_{\rm H_2}$ only. The factor of $1/2$ is due to incident radiation entering from one side only. 

The dust optical depth is related to the total column density by $\tau_d \equiv \sigma_g N$, where 
\begin{equation}
    \sigma_g=1.9\times 10^{-21} Z^{\prime}_d \;\rm{cm}^{-2}
\end{equation}
is the LW band dust-absorption cross-section \citep{Sternberg2014}. We adopt the \citep{Draine1996} expression for  $f_{\rm{sh}}$ in cool gas with temperature $\lesssim500\;\rm{K}$
\begin{align}
    f_{\rm{sh}}({N_{\rm{H}_2}})=&\frac{0.965}{(1+x/b_5)^2} + \frac{0.035}{\left(1+x\right)^{0.5}} \notag \\& \times \mathrm{exp}{[}-8.5\times 10^{-4}(1+x)^{0.5}{]},
\end{align}
where $x\equiv N_{\rm{H}_2}/(5\times 10^{14}\;\rm{cm}^{-2})$ and $b_5\equiv b/(10^{5}\;\rm{cm}\;\rm{s}^{-1})$ is the normalized absorption-line Doppler parameter. 
H$_2$ can also be destroyed by CR ionization and direct dissociation, but we choose to ignore this potential effect due to the lack of constraints in our FIRE-2  simulations on the CR flux \citep[see][ for cases in which CRs contribute to H$_2$ destruction]{Sternberg2021,Sternberg2024}. In addition, H$_2$ can be collisionally dissociated at high temperatures. In Appendix A we show that these destruction mechanisms are sub-dominant compared to photodissociation under reasonable assumptions.

\subsubsection{Subgrid Cloud Model}

In our subgrid model we consider finite 1D clouds with total column densities, $N_{\rm tot}$, that are irradiated from both sides. Cloud depth is parameterized by the gas column $N$ measured from one side, ranging from 0 to $N_{\rm tot}$. We present a detailed description in Appendix B. 

At any cloud depth we can write a formation-destruction equation for H$_2$ of the form
\begin{equation}
\label{eq: form-dest}
    R_d \, n\, n_{\ehiv{}{}}=D_{\rm LW}\, n_{\rm{H}_2},
\end{equation}
where $n$, $n_{\ehiv{}{}}$, and $n_{\rm H _2}$ are the total, atomic, and molecular number densities. By definition, it also holds that 
\begin{equation}
2n_{\rm H _2}+n_{\ehiv{}{}}=n,    
\end{equation}
where $n$ is the total (atomic+molecular) hydrogen nucleus number density of the gas, assumed to be a constant. The quantities $n_{\rm H_2}$, $n_{\ehiv}$, and $D_{\rm LW}$ all depend on the location in the cloud (as parameterized by $N$). Our goal is to find a global solution $n_{\rm H_2}(N)$, given a set of model parameters $\left( n,I_{\rm LW},Z^{\prime}_d,N_{\rm tot} \right)$. We compute the relative abundance of H and H$_2$ at each location in the cloud, which in turn gives us the total H$_2$ column $N_{\rm H_2,tot}$ associated with the cloud.

We treat every gas cell in the simulation as a 1D cloud with constant hydrogen nucleus density $n$, and a finite depth given by its total hydrogen nucleus column density $N_{\rm tot}$. The incident LW band flux is parameterized by $I_{\rm LW}$ as determined by our \skirt{} computations, and is assumed to enter at $N=0$ and $N=N_{\rm tot}$. 

Our model is iterative. For the first iteration, we assume radiation penetrating from one side only. We use eq. \ref{eq: form-dest} to find the H and H$_2$ abundances at $N=0$. We then take steps in $N$, adjusting $D_{\rm LW}$ by integrating the H$_2$ density and plugging the resulting $N_{\rm H_2}$ into eq. \ref{eq: D and f_sh}. Once we have reached $N=N_{\rm tot}$, the iteration is complete. For the following iterations, we assume radiation is penetrating from both sides, and alternate between integrating from each side. We continue to update $D_{\rm LW}$ according to eq. \ref{eq: D and f_sh} with each step while also considering the contribution from the opposite side (i.e., from the previous iteration).

We continue to iterate until the mean relative change between two consecutive iterations in the H and H$_2$ abundance drops below $0.1\%$. At this point, we obtain our result for $N_{\rm H_2,tot}$ by integrating the H$_2$ abundance across the cloud.

\subsection{Application to The FIRE Simulations}
\label{sec: subgrid application}

\begin{figure}
	
	\centering
	\includegraphics[width=1\columnwidth]{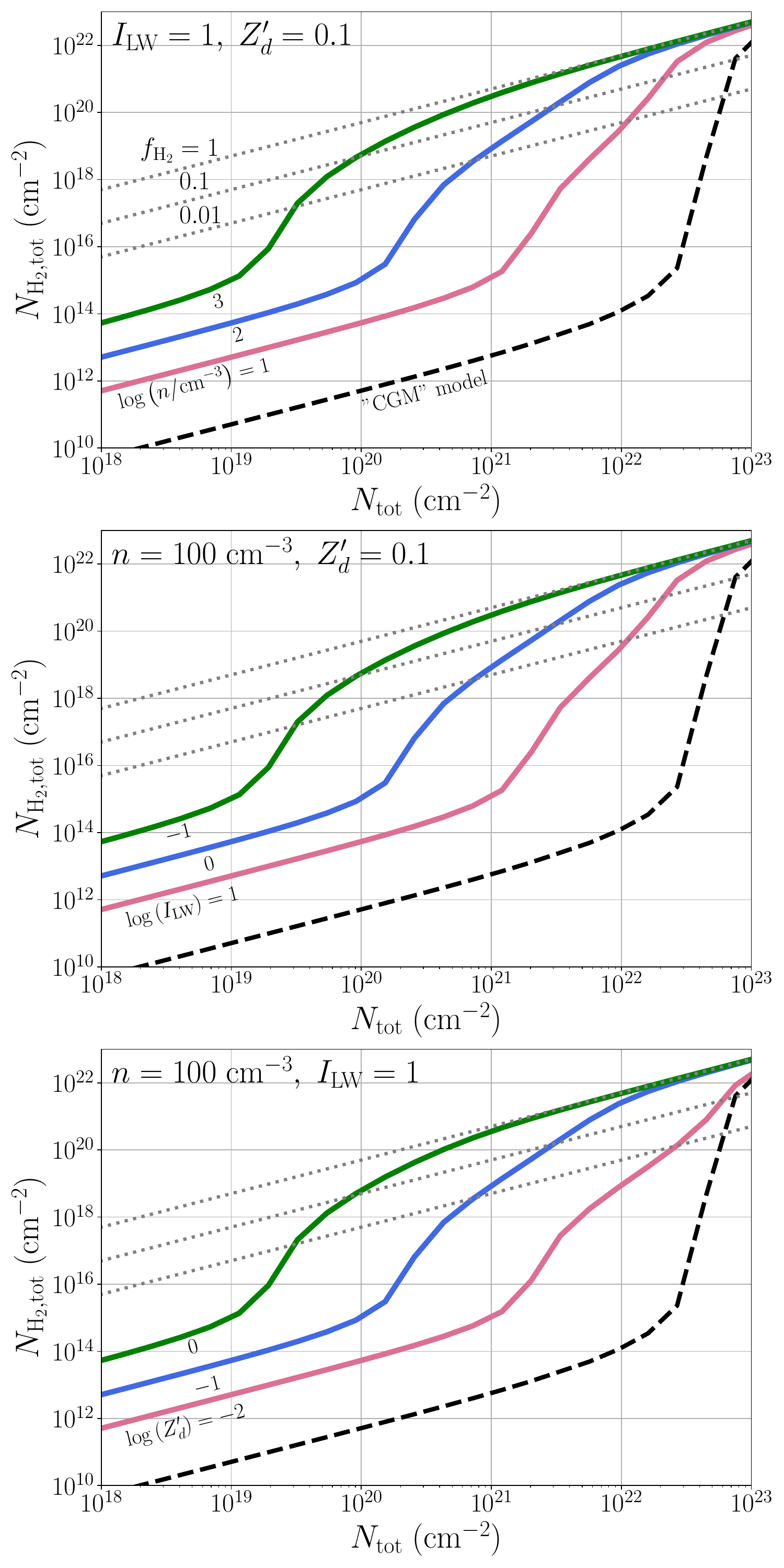} 
	\caption{
Total H$_2$ column density, $N_{\rm{H_2,tot}}$, as a function of total (atomic+molecular) hydrogen column density, $N_{\rm tot}$, for our two-sided 1D models, for various densities $n$ (top panel), LW band fluxes $I_{\rm LW}$ (middle panel), and dust abundances $Z^{\prime}_d$ (bottom panel). Colored curves show variations on our fiducial ISM model with $n=100\ \rm cm^{-3}$, $I_{\rm LW}=1$, and $Z^{\prime}_d=0.1$. Black dashed lines show our representative high-$z$ CGM model with $n=1\ \rm cm^{-3}$, $I_{\rm LW}=10$, and $Z^{\prime}_d=0.1$. Grey dotted lines indicate constant $f_{\rm H_2}\equiv 2N_{\rm H _{2,tot}}/N_{\rm tot}$.
		}
		\label{fig: H2 subgrid}
\end{figure}

We precompute and tabulate our sub-grid model results in a $50\times50\times50\times25$ lookup table. The tabulated parameters are listed in Table \ref{table: tabulated sub-grid}. We present a sample of our tabulated models in Figure \ref{fig: H2 subgrid}. We use a fiducial parameter choice of $n=100\ \rm cm^{-3}$, $I_{\rm LW}=1$, and $Z^{\prime}_{d}=0.1$ (representing a cold, neutral, and low-metallicity ISM), and increase or decrease one parameter at a time by a factor of 10. We also present a model with input parameters $n=1\ \rm cm^{-3}$, $I_{\rm LW}=10$, and $Z^{\prime}_{d}=0.1$. Motivated by the results presented in Section \ref{sec: maps}, we choose these parameters to represent conditions in the high-$z$ CGM in FIRE. We observe the well established behavior of the \hiv-to-H$_2$ transition, where a steep increase in $N_{\rm H _2}$ corresponds to a transition to a fully shielded molecular gas. The transition point is pushed to lower column densities as the density or metallicity increases, or as the LW band flux decreases. Density affects the H$_2$ column by increasing the formation rate. The dust abundance $Z_d ^{\prime}$ comes into play both in the H$_2$ formation and in the dust attenuation of LW band radiation. Finally, increasing $I_{\rm LW}$ increases H$_2$ destruction and requires a larger total column for shielding to be sufficient for conversion of the gas to fully molecular. 

The green curve in the bottom panel of Figure \ref{fig: H2 subgrid} can be considered a representative Galactic ISM model, with values typical of the cold neutral medium in the solar neighbourhood \citep{Wolfire2003}. By construction, all of the presented ISM models have higher H$_2$ columns than the CGM model, which has lower density and higher $I_{\rm LW}$. In addition, far-UV shielding is unimportant except at very high column densities $\gtrsim 3 \times 10^{22} \ \rm cm^{-2}$.

\begin{table}
\centering
\hspace*{-2cm}
\begin{threeparttable}

\centering 
\begin{tabular}{l l l l l}
\hline \hline 
\\ [-1.5ex] Parameter & $\log\left(N_{\rm tot}/\rm{cm}^{-2}\right)$ & $\log\left(n/\rm{cm}^{-3}\right)$ & $\log I_{\rm{LW}}$ & $\log Z^{\prime}_d$ \\ [0.5ex] 
\hline
\\[-2.5ex]Range & $(13,24)$ & $(-5,4)$ & $(-1,4)$ & $(-4,1)$
\\[0.5ex]Grid Points & $50$ & $50$ & $25$ & $50$
\\[0.5ex]
\hline
\end{tabular}

\end{threeparttable}
\caption{Grid points (with logarithmic spacing) for tabulation of the H$_2$ sub-grid model.}
\label{table: tabulated sub-grid}

\end{table}

As discussed in Section \ref{subsec: H2 model}, our model requires that we specify the atomic+molecular column density $N_{\rm tot}$, atomic+molecular hydrogen nucleus density $n$, LW band flux parameterized by $I_{\rm{LW}}$, and dust abundance $Z^{\prime}_d$. The metallicity and density are directly modelled by FIRE, and we can extract them from the simulation snapshots. To calculate $N_{\rm tot}$, we extract the \hi{} mass of the gas cell and divide it by the particle density in order to get an effective total length $\ell$ for the neutral gas. $N$ is then set to be $\ell\times n$. The \hi{} column from the simulation is now assumed to be the total (atomic+molecular) column, with the relative contributions of \hi{} and H$_2$ determined by our sub-grid model. 

\begin{figure}
	
	\centering
    \includegraphics[width=1\columnwidth]{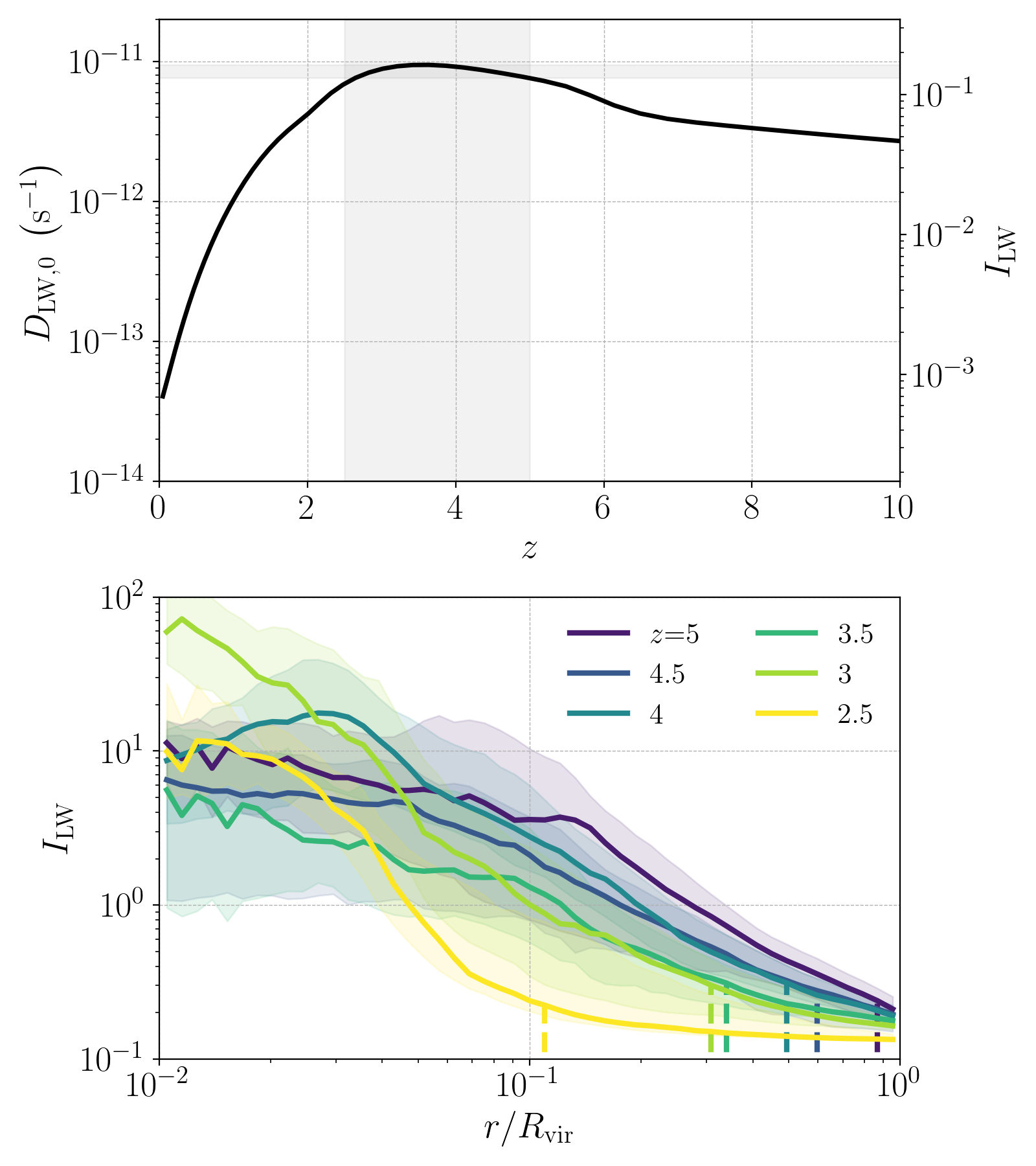} 
	\caption{
Top panel: H$_2$ photodissociation rate for a \citet{Haardt2012} UV background spectrum as a function of redshift. The right $y$-axis shows the value normalized to the solar neighborhood \citet{Draine1978} field. The shaded regions show the redshift range we study and the corresponding range in $I_{\rm LW}$ of 0.13-0.16 associated with the metagalactic field. Bottom panel: radial median profiles of $I_{\rm LW}$ for each redshift. Shaded regions show the 25-75\% range and vertical dashed lines show the radial distance at which the median contribution from the galaxy and the metagalactic radiation field are equal.
}
		\label{fig: metagalactic vs z}
\end{figure}

The LW band flux is determined using the output of the \skirt{} radiative transfer calculation, where we add a floor value by integrating a redshift-dependent \cite{Haardt2012} spectrum over the LW energy band\footnote{We compute $I_{\rm LW}=7.0\times 10^{-4}$ for a \citet{Haardt2012} spectrum at $z=0$. Integrating instead over $6-13.6$ eV, we obtain $I_{\rm UV}=4.5\times10^{-3}$, consistent with \citet{Sternberg2002}.}. In the top panel of Figure \ref{fig: metagalactic vs z} we show the redshift dependence of the metagalactic floor value that we apply to $D_{0,\rm LW}$. For the redshift range we investigate, we find that $I_{\rm LW}$ spans the range 0.13-0.16, peaking at $z=3.64$. In the bottom panel of Figure \ref{fig: metagalactic vs z} we plot radial profiles of $I_{\rm LW}$, computed by taking the median value across all halos at a given redshift. The radial distance at which the metagalactic background dominates the local LW band flux, denoted by vertical dashed lines, is 0.86, 0.60, 0.49, 0.34, 0.30, and 0.11 $R_{\rm vir}$ for $z$ going from 5 to 2.5. 

We note that the floor value of $I_{\rm LW}$ we compute is lower than the typical value of 1 in the Galaxy (and higher in the vicinity of OB stars). As we show in the following sections, the contribution to $I_{\rm LW}$ from stars in the galaxy combined with the low typical densities found in the CGM lead to ratios $I_{\rm LW}/n$ that are higher in the high-$z$ CGM compared with the ISM in the Galaxy. leading to lower H$_2$ fractions.

Using these parameters as input, we obtain a molecular column density $N_{\rm{H}_2,\rm tot}$ for each particle in the simulation volume by linearly interpolating our model grid.

\section{Results}
\label{sec: results}

\subsection{Sample Snapshot}
\label{sec: maps}
\begin{figure*}
	
	\centering
	\includegraphics[trim={0 0 0 2cm},clip,width=2\columnwidth]{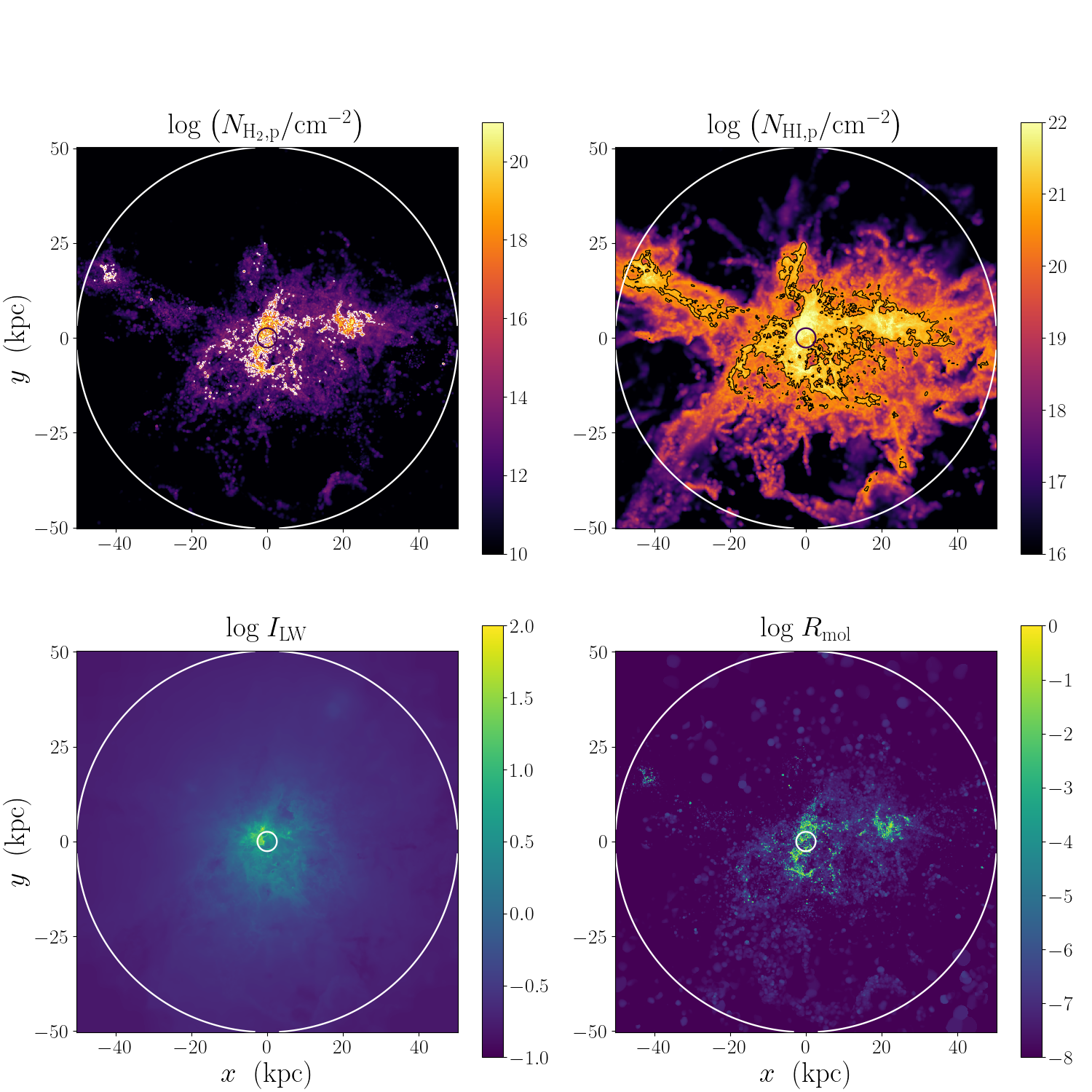 } 
	\caption{
2D maps of H$_2$ column density (top left), \hi{} column density (top left), mass-weighted average $I_{\rm LW}$, and $f_{\rm H _2}=N_{\rm H _2}/N_{\ehiv}$, for halo A4 at $z=4$ (halo mass $5\times10^{11}\ M_{\odot}$, stellar mass $0.8\times10^{10} \ M_{\odot}$, virial radius $50\ \rm kpc$, and stellar half mass radius $2.3\ \rm kpc$). White contours show $N_{\rm H_2,detec}=2\times 10^{14}\;\rm cm ^{-2}$, black contours show $N_{\ehiv,\rm DLA}=2\times10^{21}\;\rm cm ^{-2}$, the inner circle is the stellar half-mass radius, and the outer circle is the halo virial radius. 
		}
		\label{fig: A4 map}
\end{figure*}

We start by presenting a sample of our results by applying our sub-grid model to a single simulation snapshot of halo A4 at $z=4$ (halo mass $5\times10^{11}\ M_{\odot}$, stellar mass $0.8\times10^{10} \ M_{\odot}$, virial radius $50\ \rm kpc$, and stellar half mass radius $2.3\ \rm kpc$). In Figure \ref{fig: A4 map} we plot the map of \hi{} and H$_2$ column densities projected along the line of sight $N_{\ehiv,\rm p}$ and $N_{\rm H _2,p}$ for impact parameters within a $50\times 50$ kpc region. We also plot the ratio  $R_{\rm mol}\equiv N_{\rm H_2,p}/N_{\ehiv{},\rm p}$, and the mass-weighted average of $I_{\rm{LW}}$. 

As demonstrated in \citetalias{Stern2021}, DLA sightlines dominate the \hi{} map up to a significant fraction of the virial radius. The H$_2$ column density map displays a similar morphology, albeit with lower values and with a larger dynamical range. This is intuitively understood by the fact that H$_2$ is dissociated by LW band radiation that more readily escapes the central galaxies compared to ionizing radiation. Thus, $R_{\rm mol}$ remains vanishingly small even in cells that are entirely in the form of \hi{}. To discuss the detectabile sightlines in H$_2$ absoprtion, we adopt $N_{\rm H_2,detec}=2\times 10^{14} \ \rm cm^{-2}$ as a benchmark detection threshold for the observations we will compare with \citep{Ledoux2003,Noterdaeme2008,Noterdaeme2015}. The extent of detectable H$_2$ (i.e., with $N_{\rm H _2,p}>N_{\rm H_2,detec}$) is visibly larger than the stellar half-mass radius $R_{\star}$, but not as extended as the coverage of DLA sightlines. The distribution of $I_{\rm{LW}}$ shows a quasi-radial distribution, declining from a central peak of $\sim 100$ to the floor value of $\sim0.1$ set by the metagalactic radiation field at $z=4$. Within and around the stellar half-mass radius, we find sightlines with $R_{\rm{mol}}\sim 0.1$. At larger distances, $R_{\rm mol }$ decreases by orders of magnitude and the CGM mass is dominated by \hi{} (or \hii{}). We note that we may be under-predicting $R_{\rm mol}$ in the ISM of the central FIRE-2 galaxies, due to a combination of under-resolved dense gas and lack of mutual LW band shielding between different gas cells in the simulation.

\begin{figure}
	
	\centering
	\includegraphics[width=1\columnwidth]{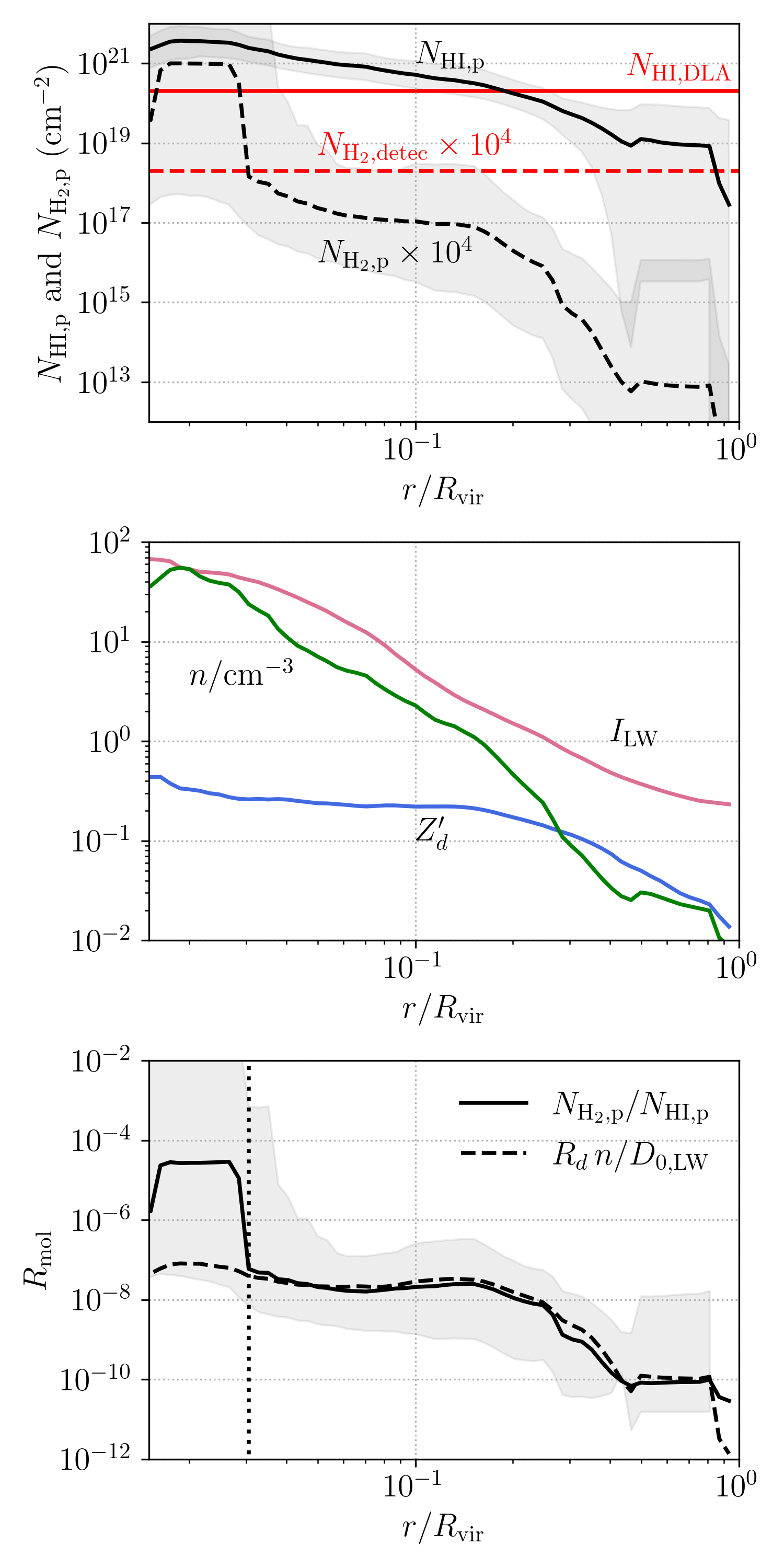} 
	\caption{
Radial profiles of different quantities for halo A4 at $z=4$. Top panel: projected \hi{} and H$_2$ column density profiles. Red solid and dashed lines show $N_{\rm \ehiv{},DLA}=2\times10^{21}\rm cm^{-2}$ and $N_{\rm H_2,detec}=10^{14} \ \rm cm^{-2}$, respectively. Middle panel: radial profiles of normalized LW band flux $I_{\rm{LW}}$, dust metallicity $Z_d^{\prime}$, and atomic+molecular hydrogen density $n$. Bottom panel: $R_{\rm mol}$ as is computed by our model (solid black) and using the optically thin approximation described in Section \ref{sec: maps} (dashed black). The vertical dotted line marks radial distance at which the dust optical depth $\tau_d=\left(N_{\ehiv{},\rm p}+2N_{\rm H_2 ,p}\right)\,\sigma_{\rm g}$ drops below 1, beyond which $R_{\rm mol}$ computed from our model agrees with the optically thin approximation. Shaded regions show the interquartile range of the corresponding curves.
		}
		\label{fig: analytic approx}
\end{figure}

We can understand the extent of the radial $N_{\rm{H}_2}$ profile as being set by $N_{\ehiv{}}$ simply scaled by $R_{\rm mol}$. The latter is, in turn, set by the ratio of the formation rate $R_d\,n$ and the destruction rate $D_{\rm LW}=D_{0,\rm LW}\,f_{\rm sh}\,e^{-\tau _d}$. For optically thin conditions, i.e., when neither dust nor H$_2$ shielding are important for determining the H$_2$ abundance, we can ignore the shielding terms and assume $D_{\rm LW}=D_{0,\rm LW}$. We therefore obtain
\begin{equation}
    R_{\rm mol}\left(r\right)=1.03\times 10^{-6} \frac{Z^{\prime}_d \left(r\right) n\left(r\right)}{I_{\rm LW}\left(r\right)}.
\end{equation}
We demonstrate the validity of this expression in Figure \ref{fig: analytic approx}, where we apply it to halo A4 at $z=4$. We construct radial profiles by binning the 2D maps according to galactocentric distance and taking the median of each quantity within every bin. We also show the interquartile ranges of $N_{\ehiv}$, $N_{\rm H_2}$, and $R_{\rm mol}$ as shaded regions about the corresponding curves. For the optically thin approximation, we use the median values for $I_{\rm LW}$, $n$, and $Z^{\prime}_d$ from our profiles. While the variation at a given radius is significant, we find that the median radial profile of $R_{\rm mol}$ is well described by the optically thin approximation over a large span of galactocentric distances. We also compute the radial profile of the LW band dust optical depth $\tau_{d}$. We approximate it as $\left(N_{\ehiv{},\rm p}+2N_{\rm H_2,p}\right)\,\sigma_{\rm g}$, where $\sigma_{\rm g}$ is computed using the radial profile of $Z_d ^{\prime}$, and mark the distance at which $\tau_d$ drops below 1. We find that this distance corresponds to the point from which the two curves agree.

\subsection{Redshift Dependence}

\begin{figure}
	
	\centering
	\includegraphics[width=\columnwidth]{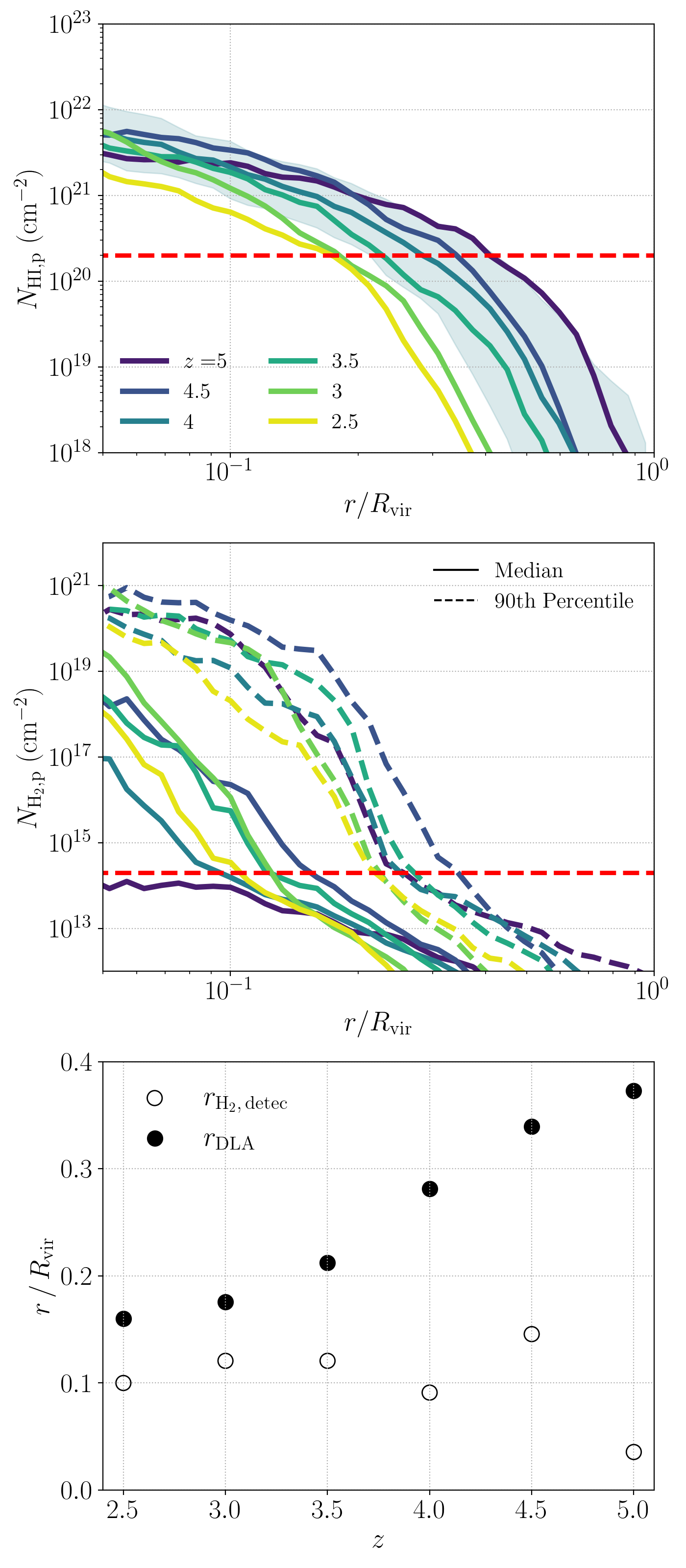} 
	\caption{
Top: radial median \hi{} column density profiles using all halos at a given redshift. The shaded region shows that interquartile range for the $z=4$ curve. Middle: as top, but for H$_2$ column density, dashed curves showing the 90th percentile. The red dashed lines correspond to the $2\times10^{21}$ (top) and $2\times10^{14}\;\rm cm^{-2}$ (middle). Bottom: the normalized radius at which the radial H$_2$ (\hi{}) profiles drop below their respective thresholds. The extent of \hi{}, as represented by $r_{\rm DLA}$ clearly increases with redshift, while the extent of detectable H$_2$ does not show a clear redshift dependence.
		}
		\label{fig: 1D prof}
\end{figure}

To investigate the redshift dependence of the H$_2$ CGM content, we construct radial profiles of the projected column densities $N_{\ehiv{}, \rm p}$ and $N_{\rm H_2,p}$. We do so by binning pixels in our maps according to their normalized impact parameter $r/R_{\rm vir}$, where $r$ is the projected distance to the halo center. We then compute the median and 90th percentile of $N_{\ehiv{},\rm p}$ and $N_{\rm H_2,\rm p}$ over all snapshots and all halos in each radial bin, and plot the results in Figure \ref{fig: 1D prof}.

The top panel of Figure \ref{fig: 1D prof} shows the median of \hi{} column density profile computed for each halo at a given redshift, with the horizontal dashed line marking the DLA threshold of $2\times 10^{21}\;\rm cm^{-2}$.  We reproduce the result of \citetalias{Stern2021}, where the \hi{} column densities at large radii are higher at higher redshifts. This is due to the CGM at higher redshift displaying larger densities, which in turn contribute to both cooling and shielding from ionizing radiation. The bottom panel of Figure \ref{fig: 1D prof} shows the normalized radius at which the \hi{} column density drops below the DLA threshold value, denoted as $r_{\rm DLA}$, demonstrating the increase in the extent of DLAs.

The middle panel of Figure \ref{fig: 1D prof} shows the averaged median and 90th percentile profiles of the H$_2$ column density. The horizontal dashed line marks our adopted detection threshold of $N_{\rm H _2}=2\times10^{14}\;\rm cm^{-2}$. We denote the distance at which the median H$_2$ column density profile drops below this threshold as $r_{\rm H_2,detec}$. Unlike the \hi{} profiles, the H$_2$ profiles do not show an obvious redshift trend, nor do they reach significant fractions of the virial radius.

\begin{figure}
	
	\centering
	\includegraphics[width=1\columnwidth]{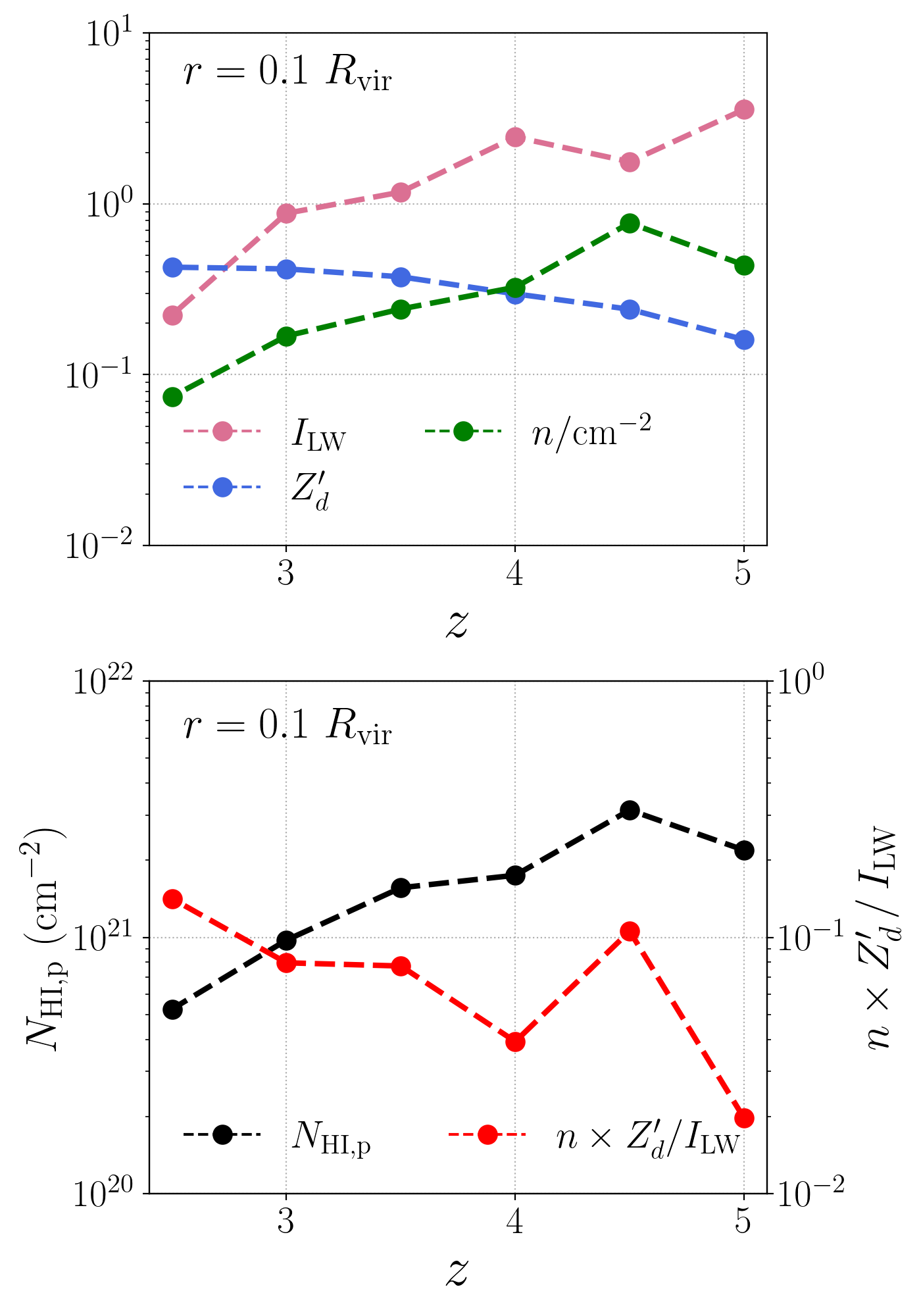} 
	\caption{
Top: redshift dependence of average $I_{\rm LW}$, $n$, and $Z_{d}^{\prime}$, computed at $r=0.1\,R_{\rm vir}$, mass-weighted along line of sight
 and averaged over all halos at a given redshift. Bottom: same as top but for $N_{\ehiv}$ and the product $n\times Z_d ^{\prime} / I_{\rm LW}$. The latter is proportional to $N_{\rm H _2,p}/N_{\ehiv,\rm p}$ under optically thin conditions.
		}
		\label{fig: 1D prof explainer}
\end{figure}

One might expect that the redshift trend in $r_{\rm DLA}$ would translate to a similar trend in $r_{\rm H _2,detec}$, but this is not the case. The explanation is demonstrated in the Figure \ref{fig: 1D prof explainer}, where we plot the values of $n$, $I_{\rm LW}$, and $Z^{\prime}_d$, computed by taking the mass-weighted average along each line of sight for each halo at $r=0.1\,R_{\rm vir}$, chosen as a rough approximation for $r_{\rm H _2,detec}$. First, the density increases with redshift. Second, the metallicity decreases with redshift because the metal production is tied to the star formation history. Third, $I_{\rm LW}$ increases with redshift, due to a combination of a decrease in the physical size of $0.1\,R_{\rm vir}$, an increase in SFR density, and a decrease in LW band shielding by dust. Finally, $N_{\ehiv,\rm p}$ increases with redshift as previously discussed. The net effect of the trends in $n$, $I_{\rm LW}$, $Z^{\prime}_d$ cancels out this increase in $N_{\ehiv{},\rm p}$, as can be seen in the bottom panel of Figure \ref{fig: 1D prof explainer}, leading to the almost redshift independent $r_{\rm H _2,detec}/R_{\rm vir}$.

\section{Comparison With Observations}
\label{sec: observations}
\subsection{DLA Sample}
In this section we compare our results to observations of H$_2$ absorption in DLAs. Our primary question is whether any of these observations can be associated with neutral CGM gas.

\begin{figure*}
	\centering
    \includegraphics[width=2.2\columnwidth]{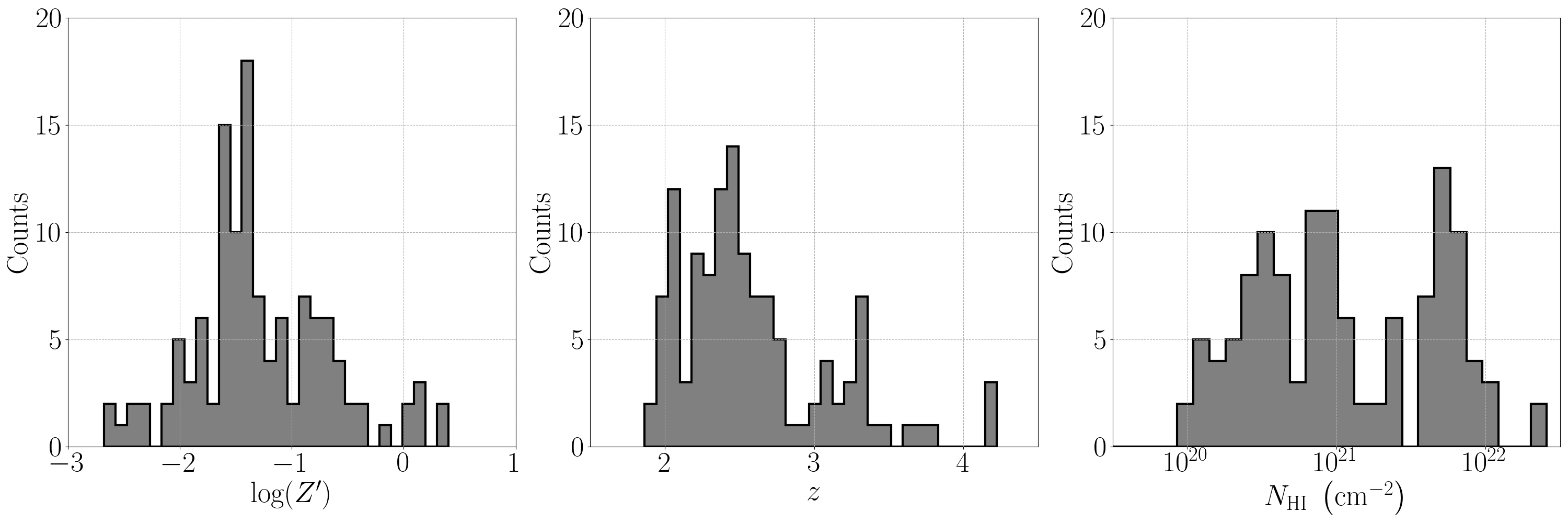}  
	\caption{
    Distributions of metallicity (left panel), redshift (middle panel), and $N_{\ehiv}$ (right panel) for the observational sample, after removing proximate DLAs.
		}
		\label{fig: sample stats}

\end{figure*}

We consider the few hundred observed DLAs investigated for H$_2$ absorption. They are characterized by some variety in the instruments used, spectral resolution (leading to different detection thresholds), and target selection criteria. The largest high resolution homogeneous effort was done by \citet{Ledoux2003} and \citet{Noterdaeme2008} using the Ultraviolet and Visual Echelle Spectrograph (UVES) on the Very Large Telescope (VLT). These observations target absorbers with $z_{\rm abs}>1.8$ and can detect H$_2$ column densities down to a characteristic detection limit of $\sim 2\times 10^{14}\ \rm cm^{-2}$. We supplement this sample with additional DLA observations performed with UVES as well as VLT/X-shooter and Keck/HIRES \citep{Ledoux2006,Noterdaeme2008b,Jorgenson2009,Jorgenson2010,Guimraes2012,Noterdaeme2015, Klimenko2016,Klimenko2020,Kulkarni2015,Albornoz2014,Balashev2010,Balashev2015,Balashev2017,Balashev2020, Noterdaeme2008b, Noterdaeme2010, Noterdaeme2017, Telikova2022, Srianand2000,Srianand2008,Ranjan2018}. We note additional observations which we do not consider in our comparison due to their inhomogeneous detection limits. These include H$_2$ detections in Sloan Digital Sky Survey (SDSS) data with a detection threshold of $\sim 10^{19}\ \rm cm^{-2}$ \citep{Balashev2014,Balashev2019MNRAS.490.2668B}, a Magellan Echellette blind search of SDSS detected DLAs with $\sim 80\%$ completeness above $10^{17.5}\ \rm cm ^{-2}$ \citep{Jorgenson2014}, and MMTO observations whith detections down to $10^{17}\ \rm cm^{-2}$ \citep{Ge1999}. We do not consider DLAs along gamma-ray burst (GRB) sightlines \citep[e.g,][]{Prochaska2009b,Ledoux2009} or proximate QSO DLAs (i.e., with a velocity offset $\Delta v<3000\ \rm km \ s^{-1}$ with respect to the background QSO) as these are generally considered to be associated with the star-forming environment from which the GRB or QSO originated. DLA observations of H$_2$ performed with the Hubble Space Telescope \citep[e.g.,][]{Muzahid2015,Boettcher2021} generally cover redshifts $<0.7$, so are also left out of our study. 
We present the distribution of metallicity, redshift, and \hi{} column density for our observational sample in Figure \ref{fig: sample stats}. $N_{\ehiv}$ shows a broad distribution covering values from $10^{20}$ to $10^{22}$ cm$^{-2}$. Most of our sample falls within the redshift range 2-3, and the characteristic metallicity is $\log Z'\approx-1.5$.
We compare our model results to the observational sample without applying any metallicity cuts. Although high-metallicity DLAs are more likely to be associated with massive halos \citepalias{Stern2021} and lead to a more fair comparison with our models, the narrow metallicity distribution of the observational sample around $Z' = 0.05$ prevents us from making any metallicity cuts to the sample without decreasing its size significantly and losing trends in the data to statistical noise.

\begin{figure*}
	\centering
    \includegraphics[width=1.8\columnwidth]{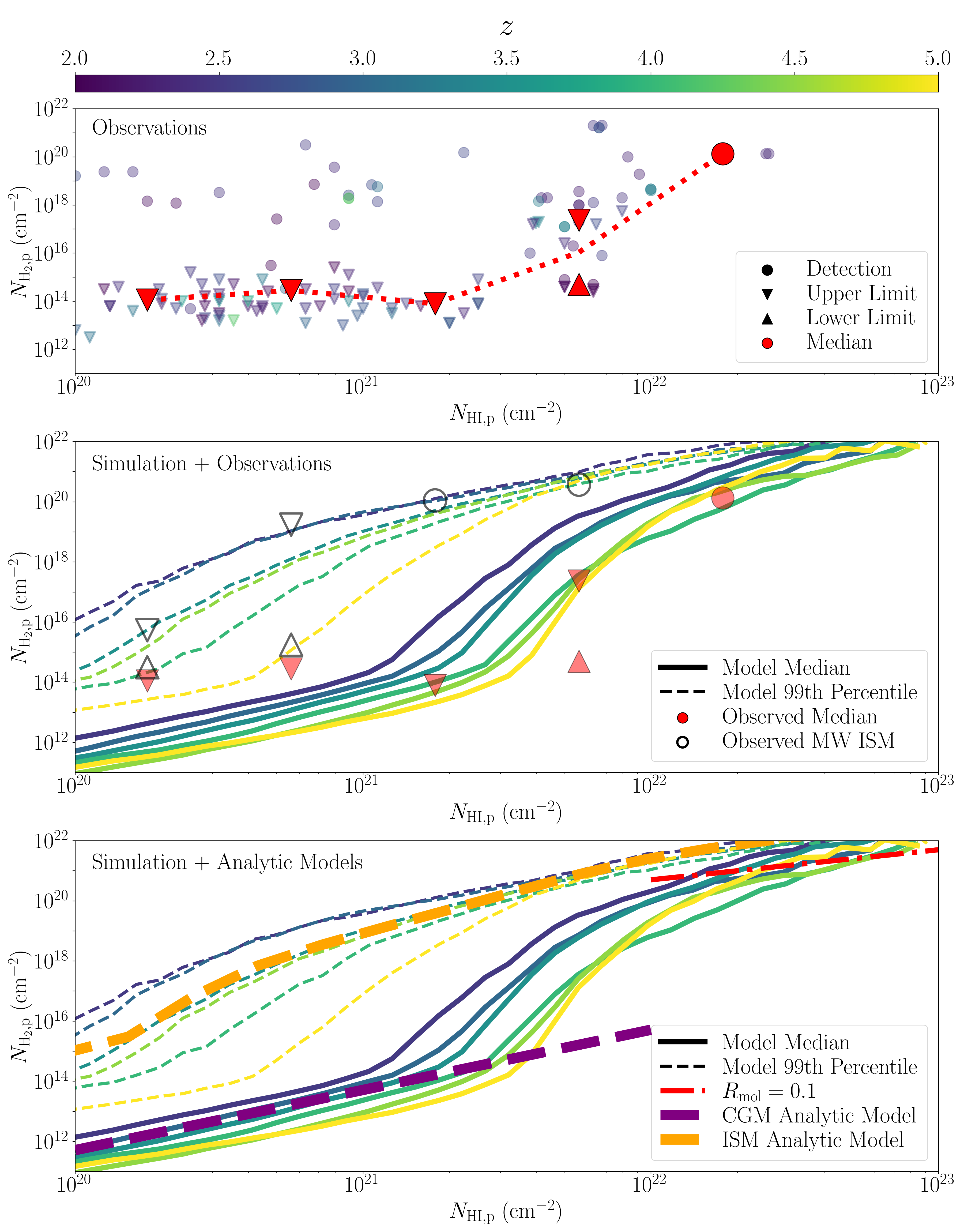}  
	\caption{
Top panel: detections (circles) and upper/lower limits (triangles) of H$_2$ vs.~\hi{} column density for the observational sample (see Section \ref{sec: observations}). 
Red symbols show observational constraints on the binned median H$_2$ column densities. 
Middle panel: median (solid lines) and 99th percentile (dashed lines) of $N_{\rm H_2,p}$ vs. $N_{\ehiv,\rm p}$ for each redshift in our model. We show again the observed medians for the high-$z$ sample and also Milky Way ISM sightlines from \citet{Savage1977}. Bottom panel: as for the middle panel, including the analytic models described in Section \ref{sec: hi to h2 fire} and a line denoting $R_{\rm mol}=0.1$ in red. 
Our analytic CGM model for the median curve assumes optically thin conditions and $n/I_{\rm LW}\propto N_{\ehiv{}}$ (see eq. \ref{eq: toy med}). The 99th percentile model assumed a 1D slab with $n=100\ \rm cm^{-3}$, $Z^{\prime}_d=0.1$, and $I_{\rm LW}=10$.
Our results reproduce the low H$_2$ columns in DLAs with lower \hi{} columns and are well described by our analytic ISM and  CGM models.
		}
		\label{fig: obs HI H2}

\end{figure*}

\subsection{\hiv{}-to-H$_2$ Transition}
\label{sec: hi to h2 fire}
In the top panel of Figure \ref{fig: obs HI H2} we show the H$_2$--\hi{} column density relation for the observational sample. The red symbols represent constraints on the medians taking into account that some of the data are upper limits. When an \hi{} column density bin contains upper limits, we first compute the median assuming the upper limits are detections. If the resulting median is larger than all upper limits, then this is the true median of the bin. Otherwise, we consider the largest upper limit in the bin to be an upper limit on the median. In addition, if there are more detections than upper limits in the bin, then we consider the lowest detection to be a lower limit on the median. The data shows an increase in the median value of $N_{\rm H_2}$ as a function of $N_{\ehiv{}}$, but we note that for most bins the median is below the detection threshold. In the middle panel of Figure \ref{fig: obs HI H2} we show the observed medians compared with our simulation results for each redshift. The latter are computed by binning all pixels according to their $N_{\ehiv , \rm p}$ value and calculating the median and 99th percentile $N_{\rm H_2, p}$ in each bin. We overplot the binned medians for Milky Way DLAs from \citet{Savage1977}. We qualitatively reproduce the observed median trend of increasing $N_{\rm H_2}$. The non-detection medians at lower column densities are also reproduced well. We over-predict $N_{\rm H_2}$ for the intermediate $\log{N_{\ehiv{}}}=21.5-22$. This could be due to our assumption of a Milky Way-like dust-to-metal ratio, which could very likely be lower at high redshift or in the CGM. In this case, we would predict lower $N_{\rm H_2}$ at a given $N_{\ehiv}$, potentially leading to a better agreement with observations. The difference between high-$z$ DLAs and the MW sightlines is clear, our model medians clearly better matching the former, while the top percentile curves are consistent with the latter.

To provide a simple interpretation of our model results we complement them with two analytic models, a CGM model and an ISM model. 
For the CGM model, we assume optically thin conditions, i.e., $D_{\rm LW}=D_{0,\rm LW}$. We also assume $Z'_d=0.1$, and $n/I_{\rm LW}=0.1\times \left( N_{\ehiv}\right/10^{20}\ \rm cm^{-2}) \ \rm cm^{-3}$. This gives rise to the relation

\begin{equation}
    N_{\rm H_2}=\left(N_{\ehiv{}}/10^{20} \rm cm^{-2} \right)^2 \times 10^{11.71}\ \rm cm^{-2}.
    \label{eq: toy med}
\end{equation}

In our ISM model, we assume a single cloud model as for our sub-grid model with $n=100\ \rm cm^{-3}$, $Z'_d=0.1$, and $I_{\rm LW}=10$, and $N_{\rm tot}=N_{\ehiv{},\rm p}$. 
By construction, applying a single cloud model assumes that the gas along the line of sight is concentrated in one physical structure in which case the projected column density fully contributes to shielding. This is unlike the optically thin CG model where we do not assume that the entire projected column density contributes to shielding, but is rather accumulated across a large volume in which different gas structures are unassociated and are not expected to mutually shield.

We present our analytic models in the bottom panel of Figure \ref{fig: obs HI H2}. Our analytic CGM model captures the qualitative shape of the median model results from FIRE-2, albeit limited to lower values of $N_{\ehiv{}}$. Our analytic ISM model captures the shape of the 99th percentile curves. The rise in the median $N_{\rm H_2,p}$ is superlinear with $N_{\ehiv, \rm p}$, as is demonstrated by our analytic CGM model. This suggests that the 3D structure of the gas is playing a role, e.g.\ due to correlations between the average density and $I_{\rm LW}$ and $N_{\rm \ehiv{},\rm p}$. Our analytic CGM model does not capture the sharp rise in the median $N_{\rm H_2,\rm p}$--$N_{\ehiv{}, \rm p}$ relation at higher $N_{\ehiv{}, \rm p}$. For the median curves, this behavior also cannot be captured by a simple 1D slab model, due to contributions from many \hi{} components along the line of sight that do not shield each other. Our analytic ISM model for the top percentile succesfully captures the highest observed values of $N_{\rm H_2}$ at a given $N_{\ehiv{}}$. This means that most of the contribution to the \hi{} and H$_2$ columns is from a single- or small number of clouds.

At low $N_{\ehiv{}, \rm p}$, our simulation results predict a trend where DLAs with a given $N_{\ehiv{}, \rm p}$ have higher $N_{\rm H _2,p}$ at lower redshift.  This trend seems to disappear at $N_{\ehiv{},\rm p}\sim 10^{22}\ \rm cm ^{-2}$. Another immediately apparent property is that the \hi{}-to-H$_2$ transition occurs at very high \hi{} columns, with the median curves crossing $R_{\rm mol }=0.1$ at $N_{\ehiv,\rm p}>10^{22}\ \rm cm^{-2}$. This is in stark contrast to the ISM in the Galaxy where gas clouds with $N_{\ehiv{}}\gtrsim 5\times 10^{20}\ \rm cm^{-2}$ already display $R_{\rm mol}\gtrsim0.1$ \citep{Savage1977, Allen2004,Gillmon2006a,Gillmon2006b,Lee2012,Lee2015,Bialy2015c,Rachford2009,Bellomi2020, VanDePutte2023}. 

We find a large vertical scatter in the observational data, and many non-detections. These indicate very low molecular fractions, mostly in DLAs with $N_{\ehiv{}}\lesssim2\times10^{21}\ \rm cm^{-2}$. At larger \hi{} columns, the upper limits fade out and are replaced by increasing numbers of actual detections. We find that the low molecular fraction in the CGM DLAs naturally explains the many H$_2$ non-detections. The combination of low metallicity, low density, and moderate-to-high LW band intensity, all lead to an $N_{\rm H _2}-N_{\ehiv{}}$ relation that is very different than observed in the ISM of the Galaxy (see Figure \ref{fig: obs HI H2}). 

At the same time, the high end of the scatter in observations is difficult to explain within our CGM framework. Indeed, there are observations that exceed the top percentile of $N_{\rm H _2, \rm p}$ in our models for a given value of $N_{\ehiv{}}$, implying that even the variability of $R_{\rm mol}$ at a given radius cannot explain them. In addition, the distribution of $N_{\rm H_2}$ at a given $N_{\ehiv}$ bin seems to be bimodal, especially at low $N_{\ehiv}$. 

A natural explanation for this is that these sightlines are simply a ``bullseye" scenario where the quasar sightline goes directly through the ISM of a galaxy \citep[see, e.g.,][]{Ranjan2018}. If this is the case, then these data would be occupying a region of $N_{\rm H _2}-N_{\ehiv{}}$ parameter space that is commonly observed in the ISM. A likely explanation for the absence of such sightlines in our simulations is that they do not correctly capture the properties of the cold ISM. In particular, the exclusion of H$_2$ self-shielding between different cells in the SKIRT model and the lack of sufficient resolution to resolve the dense molecular gas in the ISM could lead to a significant underestimate of $R_{\rm mol}$ in the ISM \citep{Seifried2018,Gurman2025}. In addition, since we do not use any large volume simulation in this work, we cannot rule out that these lines of sight might arise from less massive halos, which are more abundant than the massive halos we model. We do, however, demonstrate in our analytic ISM model that these are easy to reproduce with reasonable assumptions.

We find that our model medians over-predict $N_{\rm H_2}$ at all zredshifts when compared with observations in the $\log\left(N_{\ehiv{}} \right)=21.5-22$ bin. The median metallicity of the observed sample relative to solar is $0.05$, lower than that of $0.2-0.5$ at $0.1 \ R_{\rm vir}$ (see Figure \ref{fig: 1D prof explainer}). This difference in metallicity could be due to incomplete modeling of metal production and diffusion in FIRE-2, or the DLAs being hosted by lower mass halos. Regardless of the origin of this difference, it would lead to our models over-predicting $N_{\rm H_2}$ at a given $N_{\ehiv{}}$ due to increased H$_2$ formation on dust grains.\

Most DLAs lack an emitting counterpart. A handful of optically detected counterparts at $z\approx1-2$ \cite{Peroux2011,Peroux2011b,Peroux2016} show low impact parameters of $\approx5-10$ kpc, moderate star formation rates $\approx1-5\ M_{\odot}\ \rm yr^{-1}$, and are consistent with ISM sightlines.
On the other hand, the [C II] emitting counterparts at $z\approx 4$ presented in \citet{Neeleman2017,Neeleman2019,Neeleman2025}, with impact parameters $\approx20-60$ kpc and star formation rates $\approx10-100\ M_{\odot}\ \rm yr^{-1}$, are consistent to an extent with properties of the MassiveFIRE suite at $z=4$ \citepalias[see][for discussion]{Stern2021}. Accounting for the high-end of the $N_{\rm H_2}$ scatter, which we attribute to ISM sightlines, would likely involve simulating lower-mass halos, a representative cosmological volume, improved resolution, or some combination of the above. We discuss this further in Section \ref{sec: future}

\subsection{H$_2$ Detection Rate}

\begin{figure}
	\centering
	\includegraphics[width=1\columnwidth]{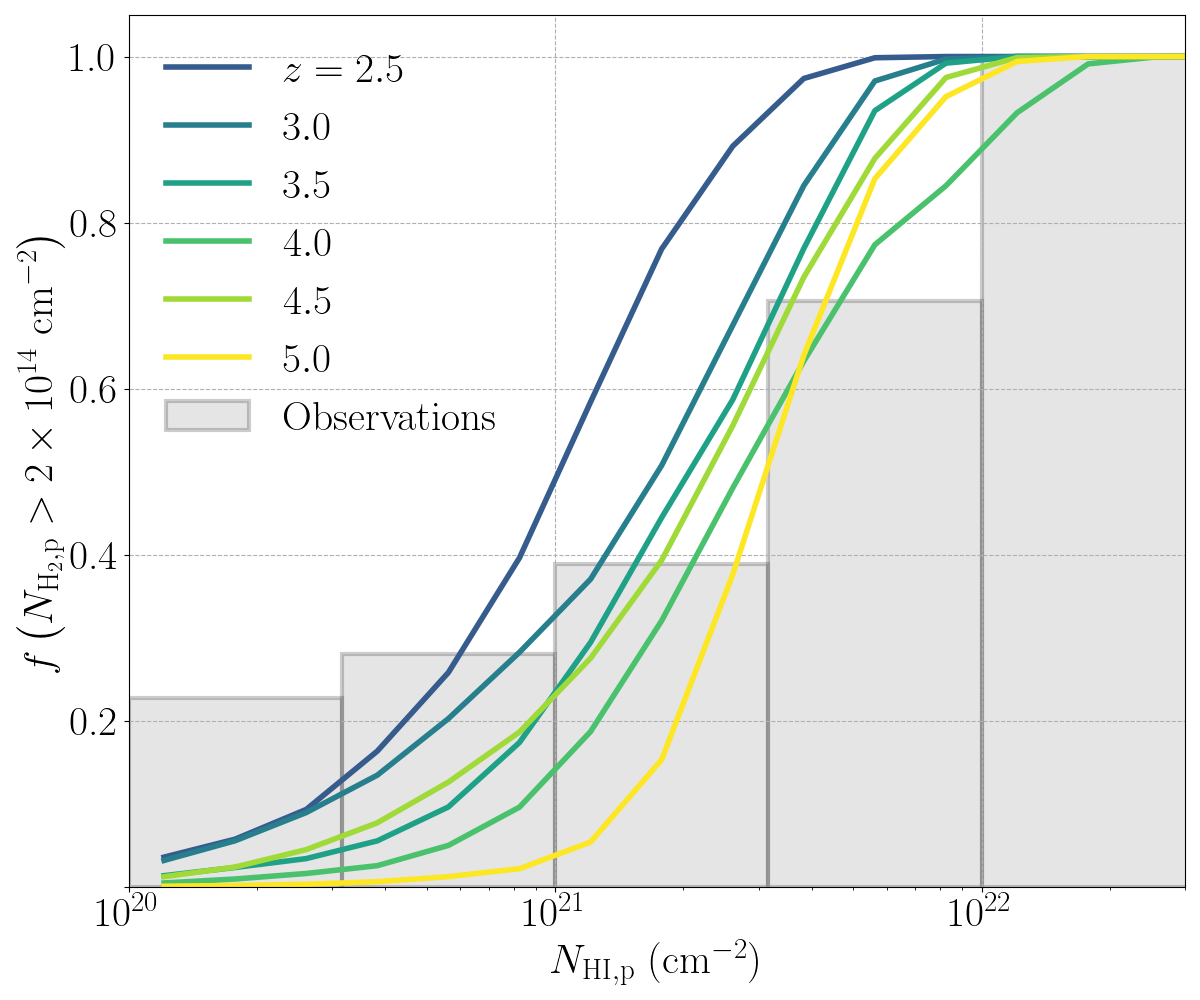}  
	\caption{
Binned detection rates as a function of \hi{} column density for the observational sample and our model.
		}
		\label{fig: obs f_detec}
\end{figure}

In Figure \ref{fig: obs f_detec} we show the detection rate in both observations and our models binned according to their \hi{} column. The predicted redshift dependence is not clear, and curves representing different redshifts seem to intersect each other. Both the data and model show a clear and predictable trend where the detection rate increases for higher values of $N_{\ehiv{}}$. The detection rate for the smallest $N_{\ehiv{}}$ bin is larger than our model result, which tends to zero at this point. This high detection rate at lower $N_{\ehiv}$, similar to the high molecular fraction discussed above, can be explained by sightlines through the ISM of galaxies. Another possible explanation, is that at the gas resolution of the FIRE-2 simulations of $\sim 10^{4}\ M_{\odot}$ is insufficient to resolve the density distribution of the CGM, even if the volume filling fraction of $T<10^{4.5}\ \rm K$ gas is converged \citep{Stern2021,Kakoly2025}. At higher resolution, the density enhancement in unresolved clumps would lead to a higher H$_2$ formation rate and abundance \citep[see, e.g., ][]{Gurman2025b}. The observed detection rate at $N_{\ehiv}\sim 6\times 10^{21}\ \rm cm^{-2}$ is lower when compared to our models. Most observations at this $N_{\ehiv{}}$ bin are at $z\sim2-2.5$, and yet fall below the $z=2.5$ curve for our simulation. This is not surprising as the constraints we find on the observed median H$_2$ in this bin in Figure \ref{fig: obs HI H2} place it lower than our model result. With 35 observations in this bin, it is unclear that sample size could be an issue here. Rather, a more likely limitation that could lead to disagreement with observations, is the lack of simulation volume representative of a typical patch of sky. The halos we analyze are rather massive and therefore represent a biased sampling of the sky. This is in contrast to DLA H$_2$ observations which are essentially an unbiased sampling of the distribution of $N_{\rm H _2}$ at a given $N_{\ehiv}$.

\subsection{Prospects for Future Models and Observations}
\label{sec: future}
While our results are instructive in explaining the low $N_{\rm H_2}$ values commonly observed in DLAs, open remaining open questions can be addressed in future work. Simulation-based models must successfully reproduce high $R_{\rm mol}$ DLAs at low $N_{\ehiv}$. As mentioned previously, several avenues must be explored. First, more precise modeling of the dense ISM including higher resolution, and, preferably, a model for sub-grid turbulence as is implemented in the FIRE-3 simulations \citep{Hopkins2023}, is likely required to correctly capture the H$_2$ fraction in the dense ISM. Second, a larger volume simulation which significantly samples the halo mass function is required to get representative statistics and detection rates, as opposed to this work which focuses and hand picked massive halos. Third, repeating this work for lower-mass and higher-resolution halos could provide insight as to the DLA-H$_2$ statistics in these more common objects. Finally, a more accurate tracking of dust evolution and dust-to-metals ratio is required. Such model exploration is already implemented in the literature \citep{Choban2022}, and as models improve and pass observational tests they can be adopted for future modeling of H$_2$ in the CGM.

At the same time, our results suggest several observational tests which could be considered as new resources become available. First, we predict that many of the H$_2$ non-detections, especially in DLAs with $N_{\ehiv}\gtrsim 10^{21}\ \rm cm^{-2}$, could actually be detected if the sensitivity of the observation were improved to $10^{13}\ \rm cm ^{-2}$. For even better sensitivity, H$_2$ should be detected at even lower $N_{\ehiv}$. Second, assuming that our simulated halos are reasonable representations of the [CII] emitting disks at $z\approx4$ detected by \citet{Neeleman2017,Neeleman2019,Neeleman2025}, our models predict that they should (on average) not be detected in H$_2$ absorption with the current detection threshold of $10^{14}\ \rm cm^{-2}$. This is due to the large impact parameters and the steep drop in $R_{\rm mol}$ with galactocentric distance in the CGM. Finally, while the sample size of the observations we compare with is significant, it is not large enough with respect to the breadth of the metallicity, redshift, and $N_{\ehiv}$ distributions it displays. A significantly larger sample size over the same parameter space would allow for a fairer comparison with models by selecting DLAs with comparable metallicities and at the appropriate redshift.

\section{Summary}
\label{sec: summary}

Recent observations and simulations suggest that the volume filling hydrogen gas in the circumgalactic medium (CGM) surrounding galaxies at redshifts $z\sim 4$ may be predominantly cool ($\sim 10^4$~K) and neutral, rather than hot and ionized as in low redshift systems. Neutral CGM gas may plausibly give rise to atomic hydrogen (\hi{}) damped Lyman-$\alpha$ absorbers (DLAs) at large offsets from the central galaxies, as is indeed indicated by observations. 
In this paper we have investigated the possibility that high-redshift neutral CGM gas may also give rise to Lyman-Werner (LW) band molecular hydrogen (H$_2$) absorptions as have occasionally been observed in DLA samples.

For this purpose we have presented computations of H$_2$ column densities in the CGM of the MassiveFIRE suite of the FIRE-2 simulations. Post processing the simulation output with our newly implemented sub-grid model for the \hiv{}-to-H$_2$ transition in conjunction with the radiative transfer code \skirt{}, we produced maps of H$_2$ column densities and compared with observations of H$_2$ absorption in DLAs.

We summarize our results as follows.

\begin{enumerate}[leftmargin=*, noitemsep]

\item The distribution of H$_2$ in the CGM shows a much lower H$_2$ fraction at a given $N_{\ehiv{}}$ compared to ISM lines of sight. The molecular fraction shows a decreasing radial dependence. This is due to a combination of a steep radial decrease in density and metallicity, and a much shallower LW band intensity decline.

\item The median H$_2$ column is detectable in absorption up to $\sim$0.1 $R_{\mathrm{vir}}$, without an apparent redshift dependence, unlike the DLA radius dependence found in \citet{Stern2021}. This is due to the redshift dependence in the formation-destruction ratio of H$_2$ compensating for the redshift trend in \hi{} column at a given (normalized) galactocentric distance.

\item H$_2$ observations of low $N_{\ehiv}$ DLAs show a low detection rate, which is inconsistent with the interpretation that they are all extragalactic ISM sightlines. We find that the low molecular fractions in the CGM of the halos we analyze provide a natural explanation for these low detection rates of H$_2$. 

\item While qualitatively explaining the low detection rates of H$_2$ and low molecular fractions in DLAs, our CGM model does not reproduce the high end of $N_{\rm H_2}$ in DLAs, likely because some of these are likely produced in the ISM of the galaxy. Future consideration of a more statistically representative sample of halos could results in a more realistic distribution, but it is also possible that DLAs are not exclusively CGM sightlines, and that additional physical interpretations of DLAs are required.
\end{enumerate}

Our models and the observations are broadly consistent with the expected properties of high-$z$ galaxy star-formation rates, CGM dust abundances, H$_2$ formation and destruction processes, and the evolution of the metalactic radiation fields. Further searches for H$_2$ absorptions in high-redshift DLAs would be most valuable.

\acknowledgements

We thank the referee for their very helpful and constructive comments. This work was supported by the German Science Foundation via DFG/DIP grant STE/1869-2 GE 625/17-1, 
by the Center for Computational Astrophysics (CCA) of the Flatiron Institute, and the Mathematics and Physical Sciences (MPS) division of the Simons Foundation, USA. JS was supported by the Israel Science Foundation (grant No.~2584/21). RKC is grateful for support from the Leverhulme Trust via the Leverhulme Early Career Fellowship. 

\bibliography{library}

\section*{Appendix A - Additional H$_2$ Formation and Destruction Mechanisms}

In this appendix we discuss the potential importance of formation and destruction processes of H$_2$ not included in our model. We consider H$_2$ formation in the gas phase via negative ion chemistry. This is a two-step process initiated by radiative attachment

\begin{equation*}
    \label{reac: rad attach}
    \rm{H}\;+\;\rm{e}^-\;\to\;\rm{H}^-\;+\;h\nu,
\end{equation*}
and followed by associative detachment

\begin{equation*}
    \label{reac: assoc detach}
    \rm{H}^-\;+\;\rm{H}\;\to\;\rm{H}_2\;+\;\rm{e}^-.
\end{equation*}
We also discuss dissociation of H$_2$ by collisions and CR ionization.

\begin{figure}
	
	\centering
	\includegraphics[width=0.7\columnwidth]{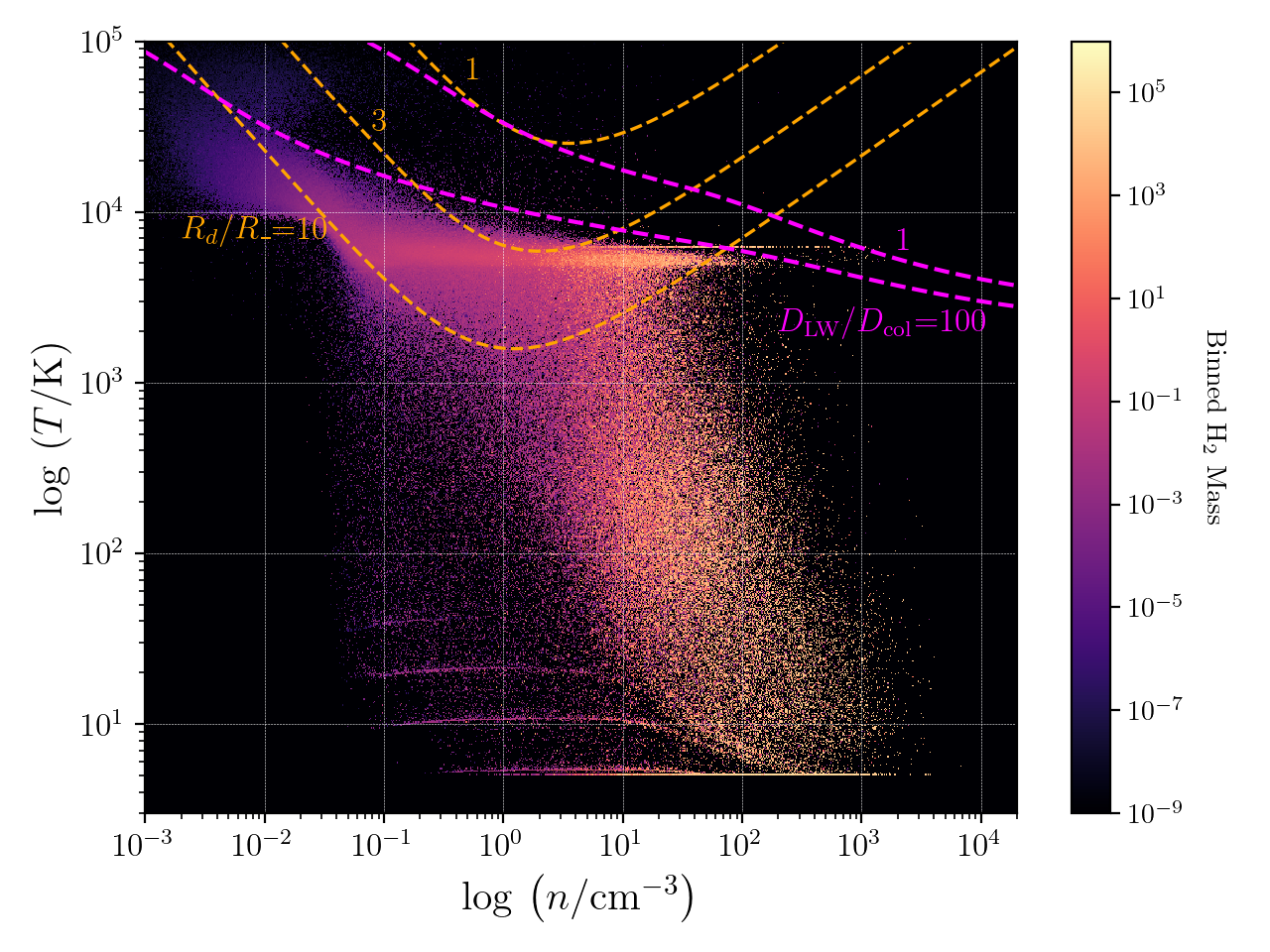} 
	\caption{
H$_2$ mass weighted histogram. Orange dashed contours delineate a constant ratio of dust-grain to gas-phase H$_2$ formation, assuming a constant CR ionization rate of $\zeta=10^{-16}$ s$^{-1}$ and solar metallicity. Pink contours delineate a constant ratio between collisional dissociation and photodissociation rates of H$_2$. 
		}
		\label{fig: 2D R D hist}
\end{figure}

In Figure \ref{fig: 2D R D hist} we present an H$_2$ mass-weighted histogram in density-temperature space, for the halo A4 at $z=4$. We overplot the contours representing $D_{\rm LW}/D_{\rm col}=100$ and 1, and $R_{d}/R_-=10$, 3, and 1, where $D_{\rm col}$ is the collisional distruction rate of H$_2$, and $R_{-}$ is H$_2$ gas phase formation rate coefficient. For $D_{\rm col}$ we use the density and temperature dependence of the collisional rate coefficient presented in \citet{Martin1996}. For $D_{\rm LW}$ we assume $I_{\rm LW}=1$ and no attenuation. For $R_{d}$ we assume $Z^{\prime}_d=1$ and $T=100\;\rm K$. For $R_{-}$ we use the expression derived in \citet{Bialy2019} and \citet{Sternberg2021}, assuming a CR H ionization rate of $10^{-16}\ \rm s^{-1}$. As expected, H$_2$ mass is concentrated at high densities, since H$_2$ formation is enhanced with respect to photodissociation at high density. 

First, we address the sub-dominance of H$_2$ destruction by CRs. As we show in Figure \ref{fig: 1D prof explainer}, our H$_2$ column density can be explained with an optically thin approximation, with the exception of a central region with area $\sim 10^{-3} \,\pi \,R_{\rm vir}^2$. This means that in most lines of sight the H$_2$ column is unshielded, and the photodissociation rate in our sub-grid cloud models can be treated as constant throughout the cloud. In Figure \ref{fig: 1D prof explainer} we see that $I_{\rm LW}\gtrsim 0.1$, and therefore $D_{\rm LW,0}\gtrsim 5.8\times 10^{-12} \ \rm s^{-1}$, in part due to the floor value set by the metagalactic background (see Figure \ref{fig: metagalactic vs z}). Thus, even at its lowest, $D_{\rm{LW,0}}$ is several orders of magnitude greater than even the highest estimates of the cosmic ray H$_2$ ionization rate in the Galaxy of $\zeta\sim 10^{-15}\ \rm s ^{-1}$ \citep{Dalgarno2006}.

In the ISM, collisional dissociation of H$_2$ is generally important in the hot ionized medium which is shock heated to $T\sim10^6\ \rm K$ by supernovae, and in shock heated molecular gas \citep{Martin1996}. The former contains no \hi{} or H$_2$ and is of no concern to us, and the latter occurs on scales that are unresolved in this work. Assuming $I_{\rm LW}=1$ we find that $D_{\rm LW}>100 \,D_{\rm col}$ for the bulk of H$_2$ mass in our model. We therefore find the omission of collisional dissociation in our model justified.

We find that gas phase formation of H$_2$ is sub-dominant at solar metallicity. For $Z^{\prime}_d=0.3$, a typical value for our $z\sim4$ halos, we find that gas phase formation is marginally important for a fraction of the gas at $n\sim1\ \rm cm^{-3}$. While this is under the assumption of $\zeta=10^{-16}\ \rm s ^{-1}$, this value is an overestimate for the CR ionization rate in the CGM, at a large distance from any star formation. As $R_{-}\propto \zeta^{-0.5}$ it is very likely that CRs do not play an important role in H$_2$ formation in the CGM in the halos that we examine. 

\label{Appendix A}

\section*{Appendix B - Subgrid Model for H$_2$ abundance}

In this appendix we describe our subgrid cloud model in full detail. The model input parameters are $\left( n,I_{\rm LW},Z^{\prime}_d,N_{\rm tot} \right)$, where $n$ is the total (atomic+molecular) hydrogen nucleus density, $I_{\rm LW}$ is the LW band intensity relative to the \citet{Draine1978} field, $Z_d ^{\prime}$ is the normalized dust abundance, and $N_{\rm tot}$ is the total column density of the cloud, and it allows us to compute the H$_2$ column for each set of parameters. We parameterize position in the cloud by the total hydrogen nucleus column density integrated up to that point, denoted by $N$. 

We start by describing our model implementation for the simple case of radiation entering the cloud from one side only. While this case has been solved analytically by \citet{Bialy2016a}, we adopt here a numerical method that can be generalized to the two-sided case and is more instructive. We begin by rearranging eq. \ref{eq: form-dest} into the form
\begin{equation}
\label{eq: cloud equation}
    \frac{n_{\rm H _2}}{n_{\ehiv{}{}}} \left(N \right)=\frac{2R_d}{D_{0,\rm LW} e^{-\tau_d} f_{\rm sh} \left(N_{\rm H _2} \right)}.
\end{equation}

We first define a discrete logarithmic grid of $k$ values $\{ N_i\}_{i=0} ^{k}$ going from $N_0=0$ to $N_k=N_{\rm tot}$, with our goal being to compute the corresponding values of $\{N_{\text{H}_2,i}\}$, and, in turn, $\{ n_{\text{H}_2 ,i} \}$. We set $k=5\times 10^3$ (i.e., $dN/N\approx0.4 \%$), and note that the model achieves convergence with respect to $k$ already for $k=100$, for most parameter choices. For convenience we will refer to $N=0$ as the left side of the cloud and $N=N_{\rm tot}$ as the right side, and denote $N_{\rm H_2,tot}=N_{\rm H_2}\left(N_{\rm tot} \right)$ and $\tau_{d,\rm tot}=\tau_{d}\left(N_{\rm tot} \right)$. Next, we set the initial condition $N_{\text{H}_2 ,0}=0$ at $N_0=0$, allowing us to solve eq. \ref{eq: cloud equation}, obtaining $n_{\ehiv{}{}, 0}$ and $n_{\rm H _2,0}$. We then solve for the subsequent indices by first computing $N_{\text{H}_2,i}$ using
\begin{equation}
    N_{\text{H}_2,i}=\int_0 ^ {N_i} \frac{n_{\mathrm{ H _2},i-1}}{n}dN^{\prime}.
\end{equation}
Finally, we use the value of $N_{\text{H} _2,i}$ to compute $f_{\rm sh}$, and subsequently $n_{\ehiv{}{}, i}$ and $n_{\text{H} _2,i}$. We iterate the process over all ${N_i}$ until we reach $N_{\rm tot}$.

To generalize this procedure for the two-sided case, we present the following modified version of eq. \ref{eq: cloud equation}

\begin{equation}
    \label{eq: 2-side cloud equation}
    \frac{n_{\rm H _2}}{n_{\ehiv{}{}}} \left(N \right)= \frac{2R_d}{D_{0,\rm LW} \left( e^{-\tau_d} f_{\rm sh} \left(N_{\rm H _2} \right)+e^{-\left({\tau}_{d,\rm tot}-\tau_d\right)} f_{\rm sh} \left(N_{\rm H_2,tot}-{N}_{\rm H _2} \right)\right)},
\end{equation}

In this expression, $N_{\rm H_2,tot}-{N}_{\rm H _2}$ and ${\tau}_{d,\rm tot}-\tau_d$ are the dust optical depth and H$_2$ column density at depth $N$, but integrated from the right side rather than the left. As such, $D_{\rm 0,LW}e^{-\left({\tau}_{d,\rm tot}-\tau_d\right)} f_{\rm sh} \left(N_{\rm H_2,tot}-{N}_{\rm H _2}\right)$ represents the radiation coming in from the right.

\begin{figure}
	
	\centering
	\includegraphics[width=0.9\columnwidth]{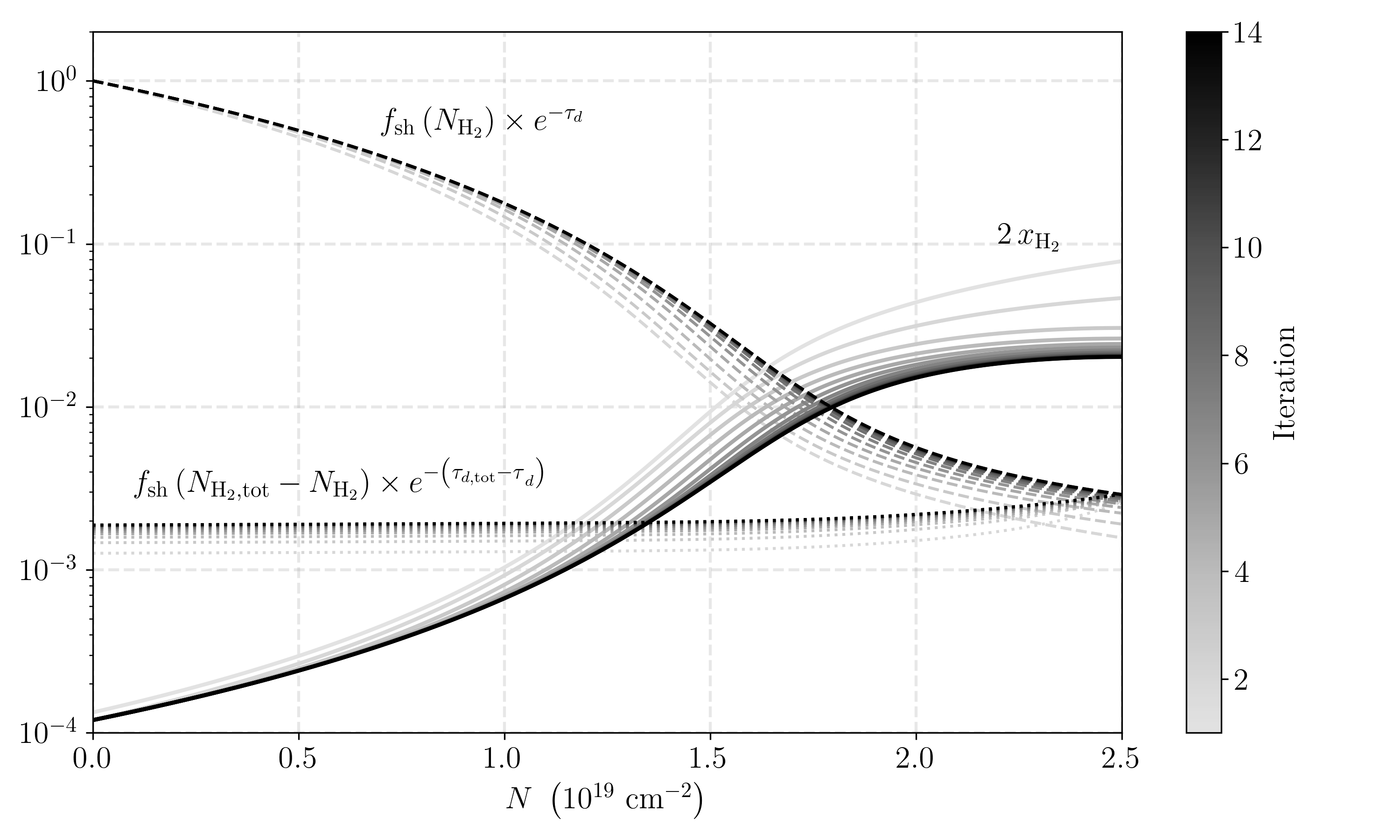} 
	\caption{
Visualization of the model convergence over 14 iterations for $I_{\rm LW}=1$, $n=100\ \rm cm^{-3}$, $Z'_d=1$, and $N_{\rm tot}=5\times 10^{19} \ \rm cm^{-2}$. Different curves show the profiles of $2\,x_{\rm H_2}$ (solid), $f_{\rm sh}\left(N_{\rm H_2}\right)e^{-\tau_d}$, and $f_{\rm sh}\left(N_{\rm H_2,tot}-N_{\rm H_2}\right)e^{-\left(\tau_{d\rm,tot}-\tau_d\right)}$ (dashed), from the different iterations indicated by the line colors.
		}
		\label{fig: model vis}
\end{figure}

For clarity of the following expressions, we now remove the subscript $i$ which we used to refer to the discrete points on our grid of discretized values of $N$, and now use the subscript $j$ to refer to iterations of the global profiles $n_{\text{H}_2,j}(N)$, $n_{\ehiv{}{},j}(N)$, and $N_{\text{H}_2,j}(N)$. The $j=0$ profiles are determined by the one-sided case as explained above. Next, we add radiation from the righthand side, giving the equation

\begin{equation}
    \frac{n_{\text{H} _2,j+1}}{n_{\ehiv{}{},j+1}} \left(N \right)= \frac{R_d}{D_{0,\rm LW} \left( e^{-\tau_{d,j}} f_{\rm sh} \left(N_{{\rm H _2},j} \right)+e^{-\left({\tau}_{d,\rm tot}-\tau_{d,j}\right)} f_{\rm sh} \left(N_{\rm H_2,tot}-{N}_{{\rm H _2},j} \right)\right)}.
\end{equation}
We then compute $n_{\text{H}_2,j+1}(N)$ and complete the step by setting
\begin{equation}
    N_{\text{H} _2,j+1}(N)=\int_0 ^N \frac{n_{\text{H}_2,j+1}}{n}dN^{\prime},
    \label{eq: NH2 int}
\end{equation}
and repeat this process iteratively until the mean relative difference between $n_{\text{H}_2,j}(N)$ and $n_{\text{H}_2,j+1}(N)$ drops below $0.1\%$. At the end of this procedure, we obtain $N_{\rm H_2}$ for the cloud by the plugging in $N=N_{\rm tot}$ in eq. \ref{eq: NH2 int}. Figure \ref{fig: model vis} demonstrates the iterative process. We plot the resulting profiles of $x_{\rm H_2}$, 
$f_{\rm sh}\left(N_{\rm H_2}\right)e^{-\tau_d}$, and $f_{\rm sh}\left(N_{\rm H_2,tot}-N_{\rm H_2}\right)e^{-\left(\tau_{d\rm,tot}-\tau_d\right)}$
for the left half of the cloud in the 14 iterations required to achieve convergence for the case $I_{\rm LW}=1$, $n=100\ \rm cm^{-3}$, and $Z'_d=1$. The iteration number is indicated by the color of the curves. We can see that the LW-band flux contribution from the left (right) decreases (increases) with increasing $N$. As is expected, 
$f_{\rm sh}\left(N_{\rm H_2}\right)e^{-\tau_d}$ = $f_{\rm sh}\left(N_{\rm H_2,tot}-N_{\rm H_2}\right)e^{-\left(\tau_{d\rm,tot}-\tau_d\right)}$
at the center of the cloud once convergence is obtained, i.e., the LW-band flux from the left and right side of the cloud are equal at the center.

\end{document}